\documentclass[aps,10pt,prb,twocolumn,superscriptaddress,showpacs,floats,floatfix,preprintnumbers,amsmath,amssymb]{revtex4-1}
\usepackage[T1]{fontenc}
\usepackage[latin9]{inputenc}
\usepackage[USenglish,american]{babel}
\usepackage{amsmath}
\usepackage{amssymb}
\usepackage{wasysym}
\usepackage{graphicx}
\usepackage{tabularx}
\usepackage{xcolor}
\usepackage{bm}          % bold math
\usepackage{pifont} 
\usepackage{dcolumn}
\usepackage{float}
\usepackage{natbib}
\newcommand{\mg}{MgB$_2$}
\newcommand{\tc}{T$_{\textrm{c}}$}
\newcommand{\ep}{$ep$}

\newcommand{\UniRoma}{Dipartimento di Fisica, Universit\`a di Roma La Sapienza, Piazzale Aldo Moro 5, I-00185 Roma, Italy}
\newcommand{\GrazPhys}{Graz University of Technology, NAWI Graz, 8010 Graz, Austria}
\newcommand{\UniCornell}{Department of Materials Science and Engineering, Cornell University, Ithaca, New York 14853, USA}

\newcommand{\UniBern}{Department of Chemistry and Biochemistry, University of Bern, Freiestrasse 3, CH-3012 Bern, Switzerland}
\begin{document}
\title{High-Temperature Conventional Superconductivity in the Boron-Carbon system:
Material Trends
}

\author{Santanu Saha}               \affiliation{\GrazPhys}
\email{santanu.saha@tugraz.at}
\author{Simone Di Cataldo}  \affiliation{\GrazPhys} \affiliation{\UniRoma}
\author{Maximilian Amsler} \affiliation{\UniCornell} \affiliation{\UniBern}
\author{Wolfgang von der Linden}    \affiliation{\GrazPhys}
\author{Lilia Boeri}                \affiliation{\UniRoma}

\date{\today}

\begin{abstract}
In this work we probe the possibility of high-temperature conventional superconductivity in the boron-carbon system, 
using {\em ab-initio} screening. A  database of 320 metastable structures with fixed composition (50$\%$/50$\%$) is 
generated with the Minima-Hopping method, and  characterized with electronic and vibrational descriptors.
Full electron-phonon calculations on sixteen {\em representative} structures allow to identify general trends in \tc\; 
across and within the four families in the energy landscape, and to construct an approximate \tc\; predictor, based
on transparently interpretable and easily computable electronic and vibrational descriptors. Based on these, we estimate 
that around 10$\%$ of all metallic structures should exhibit \tc\;'s above 30 $K$. This work is a first step towards {\em ab-initio} 
design of new high-\tc\; superconductors.
\end{abstract}

\maketitle

\section{Introduction}\label{Sec:Intro}
For more than one century, the discovery of a room temperature superconductor has 
been considered one of the "holy grails" of condensed matter physics. Already at 
the end of the 60's, N.W. Ashcroft and V.Ginzburg~\cite{ashcroft1968metallic}
predicted that under sufficiently high pressures hydrogen may be turned into an atomic
metallic phase,~\cite{Huntington1935} which would behave as a high-temperature superconductor (HTS). However,
until last year,  the pressures involved in hydrogen metallization turned out to be
prohibitive, even for the best high-pressure research labs
worldwide.~\cite{Hydrides:our_review,H:Eremets_2019,H:Loubeyre_2020}
 
In 2004, Ashcroft proposed that the metallization pressure may be sensibly reduced by
exploiting chemical  precompression of the hydrogen sublattice in H-rich compounds.
SH$_3$\cite{PhysRevLett.92.187002} , predicted in 2014 by Duan et.
al.\cite{duan2014pressure},  was experimentally found to be superconducting at 
200 GPa with a \tc\;=203 K by the Eremets' group in
2015;\cite{drozdov2015superconductivity} in
less than five years, the maximum \tc\; in high-pressure hydrides was raised up to 
265 K in LaH$_{10}$, predicted a few years before.\cite{liu2017potential,LaH10:hemley_PRL2019,drozdov2019superconductivity}

Although room-temperature superconductivity at high pressures is an impressive 
achievement by itself, practical applications of superconductivity require materials
that can operate at ambient pressure. Thus, the main focus of superconductivity research
is gradually shifting from room temperature superconductivity at high pressures to HTS
at ambient pressure.\cite{SC:Boeri_Bachelet_JPCM2019}

Proposals to realize HTS at ambient pressure based on the conventional electron-phonon
(\ep) mechanism - High-\tc\; Conventional Superconductivity (HTCS) - date back to the
early 2000's, when the MgB$_2$ discovery\cite{nagamatsu2001superconductivity}
showed that HTCS are best realized in {\em covalent metals},\cite{SC:mgb2_Pickett_PRL2001}
where the high-phonon frequencies and strong \ep\; matrix elements typical of
covalent bonds coexist with metallic behavior,
which is a prerequisite for conventional 
superconductivity.\cite{ekimov2004superconductivity,SC:diamond_Boeri_PRL2004,SC:diamond_Pickett_PRL2004,SC:Blase_PRL_2004,SC:diamond_giustino_PRL_2007}
Following this general principle, several hypothetical materials were proposed: 
notable examples are doped LiBC, hexagonal Li-B, graphane, etc~\cite{SC:Rosner_LiBC_PRL2002,SC:LIB_Kolmogorov_PRB2006,SC:graphane_savini_PRL2010}.
These are all chemical and structural analogs of \mg,
proposed on the basis of simple physical arguments, but 
without a knowledge of the underlying thermodynamics.

Only recently, the wide-spread use of modern methods for crystal structure prediction (CSP) has permitted 
to address the crucial aspect of thermodynamics
in material design.
Combined with methods for high-throughput (HT) database screening and machine 
learning (ML), CSP methods are an unprecedentedly powerful tool driving a sudden acceleration
in material discoveries in the last few years~\cite{Hydrides:our_review,ML:MLroadmap_Alberi_2018,Schleder_MLreview_2019}.
However, compared to other problems of material research, their application
to superconductivity is still at a very early
stage,~\cite{ML:Stanev_TCML_2018,Ishikawa_ML_PRB2019,ML:Eliashberg_Xie_PRB2019,ML:Hutcheon_Hydrides_arxiv2020} due to two
intrinsic problems: ($i$) for a large class of {\em unconventional} superconductors,
including the high-\tc\; cuprates, a quantitative theory of superconductivity is currently
missing ; ($ii$) for {\em conventional} superconductors where, on the other hand, \tc\;
can be predicted with remarkable accuracy, the cost of a single \tc\; calculation is too
high, to directly perform high-throughput screening of large databases of hypothetical
materials.~\cite{SC:Boeri_Bachelet_JPCM2019}

This work is part of a larger project, in which we plan to investigate superconductivity
trends across several families of conventional superconductors, to identify meaningful
screening protocols to search for promising superconducting candidates. In this paper, 
we focus on boron-carbon (BC) structures, with a 50\%-50\% composition.

First, we generate a large database of 320 distinct metastable boron/carbon structures with the minima hopping 
method (MH)~\cite{Goedecker_2004,Goedecker_2005,Amsler_2010}.
The whole set is then analyzed to identify the main structural templates characterizing the potential energy surface; 
on the basis of simple electronic and vibrational descriptors, the number of structures is progressively
narrowed down to a set of sixteen \textit{representative} structures, for which we
perform full \tc\; calculations, to identify and understand empirical trends
governing superconductivity in BC systems.

The BC system is an ideal choice for a first blind study of superconductivity, because
both boron and carbon are light elements which tend to form stiff, directional bonds, 
favorable for HCTS; furthermore, both elements exhibit a strong tendency to
polymorphism\cite{jay2019theoretical}, which ensures that the pool of structures
generated by MH will be large and diverse.
Several studies in literature have already predicted  conventional superconductivity
with sizable \tc\; in the boron-carbon system for selected compositions and structural
motifs;\cite{SC:diamond_Boeri_PRL2004,SC:diamond_Pickett_PRL2004,SC:calandra_B12_2004,SC:Calandra_BC5_PRL2008}
a series of pioneering works by Moussa and Cohen analyzed \tc\; trends 
in selected templates, using the rigid-band approximation for doping and the
rigid-muffin-tin approximation
for the \ep\; coupling,~\cite{Moussa_diamond_PRB2008,SC:Moussa_bounds_PRB2006,SC:Moussa_molfrag_PRB2008} but to our knowledge this is the first work
which exploits CSP methods to generate physically-meaningful structures and systematically investigate their superconducting properties. 

The boron-carbon phase diagram is extremely complex;
hexagonal and tetrahedral motifs, characteristic of C $sp^2$/$sp^3$ bonds,  dominate the energy landscape up to
 $\sim 1/3:2/3$ $C:B$  compositions, while more complex motifs develop for higher B concentrations, 
due to an increasing role of electron-deficient boron.~\cite{jay2019theoretical}
To limit the scope of our analysis, we decided to focus on the single 50$\%$-50$\%$ composition, where the physics and chemistry
should still be dominated by carbon, but boron is in a sufficiently high concentration to ensure that many phases will exhibit a 
pronounced metallic behavior.

This paper is organized as follows. In section \ref{sect:DB} we discuss the general
features of the whole pool of 320 structures, their classification into different
families and their salient qualities.  We also describe briefly how the sixteen
representative structures are selected for our subsequent superconductivity studies. 
In section \ref{sect:trends} we discuss the trends in \tc\; amongst different 
structures, and how they are correlated with electronic structure quantities. 
In section \ref{sect:elestruct} we discuss in greater detail the electronic, 
vibrational and superconducting properties of the structures. In section
\ref{sect:Tcformula} we show that a simple analytical expression interpolates 
the \tc\; of the {\em representative} structures, and may be used as a predictor
for superconductivity. Finally in section \ref{sect:conclusions} we 
summarize the main conclusions of our work. Appendix A contains plots of the 
electronic and phononic DOS, and Migdal-Eliashberg spectral functions
for the sixteen representative structures, while the methodology
used for this study is discussed in greater detail in the Appendix B. 
In addition convex hull has also been provided in Appendix B.

%%%%%%%%%%%%%%%%%%%%%%%%%FIG 1 -ENEVOL and TEMPLATE
\begin{figure*}[!htb]
\includegraphics[width=1.5\columnwidth,angle=0]{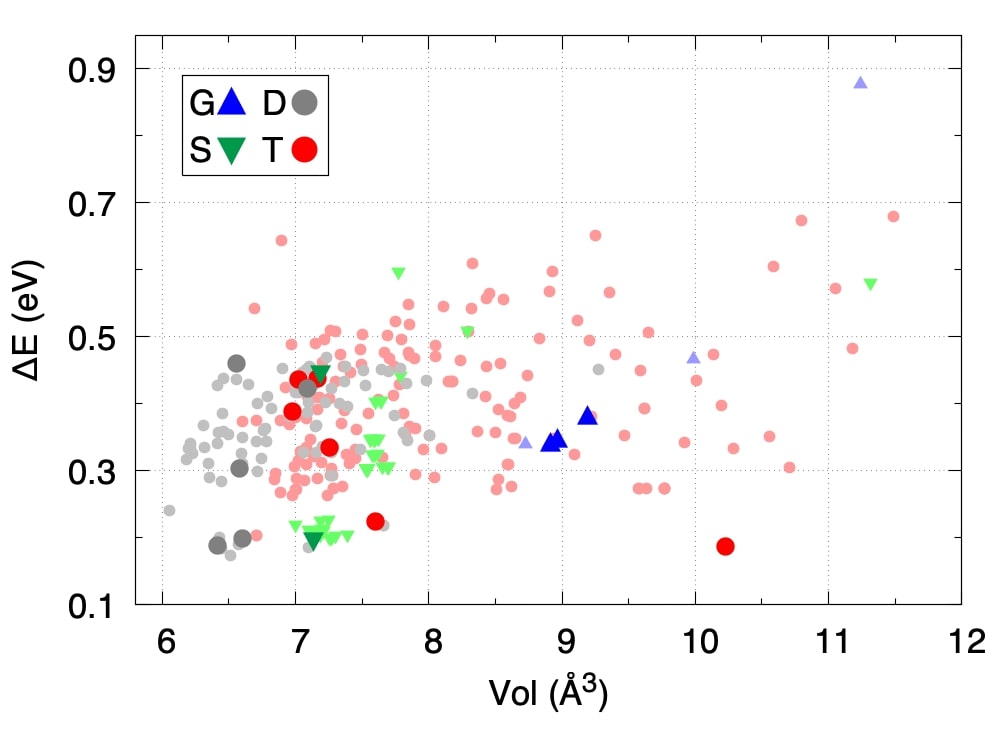}
\includegraphics[width=1.5\columnwidth,angle=0]{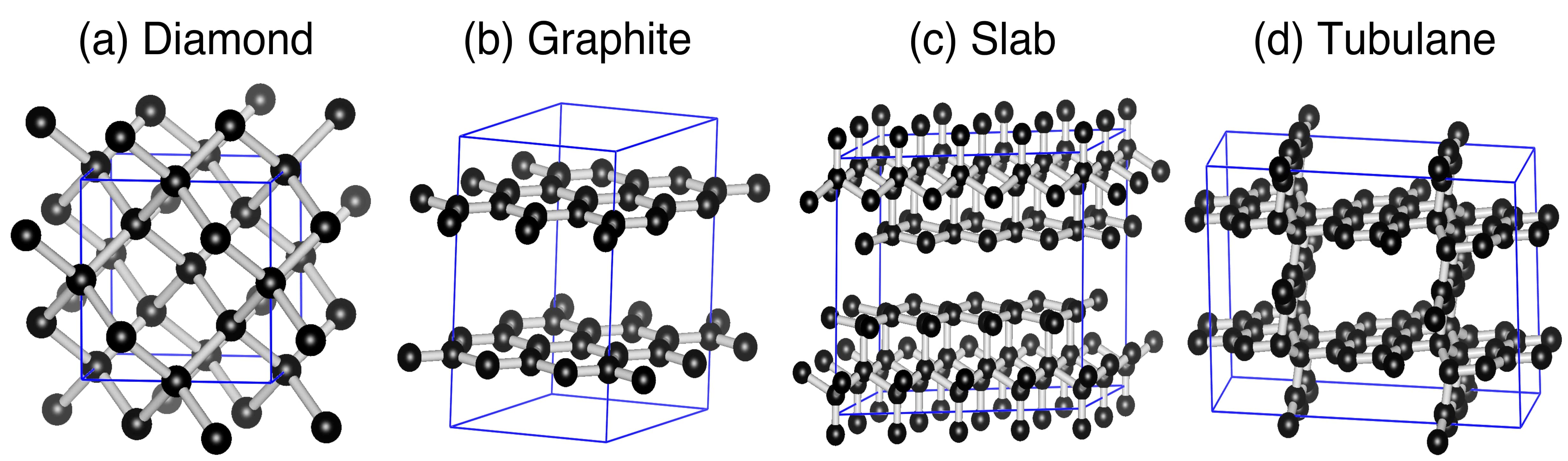}
\caption{The Volumes vs Formation Energies of all predicted BC structures are shown
in (\AA$^3$/atom) and (eV/atom) respectively. The formation energy is calculated 
w.r.t the energy of graphite and $\alpha$-Rhombohedral-B$_{12}$. The colored symbols
in the  plot indicate the family of the structure, i.e. diamond (D) structures are
represented by grey circles, graphite (G) by blue triangles, slab (S) by inverted 
green triangles and tubulane (T) by red circles. Large dark-coloured symbols indicate
the sixteen \textit{representative} structures which we selected for
further study, while the rest are shown by small light-coloured symbols. 
Typical structures of the families diamond, graphite, slab and tubulane are shown in
(a),(b), (c) and (d) respectively; black spheres represent  atoms and off-white
cylinders bonds. Considering the full convex hull, the formation enthalpy may be
uniformly shifted up or down, depending on the carbon precursor.}
\label{fig:volene}
\end{figure*}

%%%%%%%%%%%%%%%%%%%%%%%%%%%%%%%TABLE I - ALL DATA OF SELECTED B4C4
\begin{table*}[!thb]  
\centering
\begin{tabular}{cccccccccccccc}
\hline
ID & Space & Vol & \text{$\Delta$E} & N($E_F$) & $\omega_{max}$ &
$\omega_{avg}$ & $\lambda$ & $\lambda$/N($E_F$) & $\omega_{log}$ & $T_c$ & \multicolumn{3}{c}{Bond Type}\\
  & Gr. Ind. & (\AA$^3$) & (meV) & (states/eV) & (meV) & (meV) & & & (meV) & (K) & CC & BB & BC\\
\hline\hline

\multicolumn{2}{l}{Diamond} & \multicolumn{9}{c}{} \\
\hline
D01 & 164 & 6.30 & 170 & 0.15 & 160 & 79 & 0.6 & 4.2 & 72 & 21 & 
\ding{52} & \ding{55} & \ding{52} \\
D02 & 008 & 6.47 & 190 & 0.12 & 162 & 84 & 0.5 & 4.3 & 62 & 10 & \ding{52} & \ding{55} & \ding{52}\\
D03 & 051 & 6.46 & 270 & 0.16 & 152 & 92 & 0.8 & 4.9 & 72 & 35 & \ding{52} & \ding{52} & \ding{52}\\
D04 & 160 & 6.92 & 420 & 0.21 & 128 & 92 & 0.8 & 3.9 & 52 & 30 & \ding{55} & \ding{55} & \ding{52}\\
D05 & 216 & 6.56 & 440 & 0.36 & 106 & 71 & 2.3 & 6.3 & 41 & 75 & \ding{55} & \ding{55} & \ding{52}\\

\hline\hline 
\multicolumn{2}{l}{Graphite} & \multicolumn{9}{c}{} \\ 
\hline
G01 & 012 & 8.57 & 360 & 0.09 & 186 & 95 & 0.4 & 4.1 & 42 & 2 & \ding{52} & \ding{52} & \ding{52} \\
G02 & 012 & 8.63 & 360 & 0.11 & 186 & 95 & 0.4 & 3.4 & 39 & 1 & \ding{52} & \ding{52} & \ding{52}\\
G03 & 002 & 8.91 & 390 & 0.08 & 198 & 104 & 0.4 & 4.3 & 32 & 1 & \ding{52} & \ding{52} & \ding{52}\\

\hline\hline 
\multicolumn{2}{l}{Slab} & \multicolumn{9}{c}{} \\
\hline
S01 & 164 & 6.96 & 190 & 0.16 & 161 & 76 & 0.6 & 4.1 & 79 & 25 & \ding{52} & \ding{55} & \ding{52}\\
S02 & 156 & 7.07 & 440 & 0.21 & 138 & 72 & 1.1 & 5.3 & 57 & 53 & \ding{55} & \ding{55} & \ding{52}\\

\hline\hline 
\multicolumn{2}{l}{Tubulane} & \multicolumn{9}{c}{} \\
\hline
T01 & 044 & 10.02 & 150 & 0.14 & 184 & 98 & 0.3 & 1.9 & 44 & 0 & \ding{52} & \ding{55} & \ding{52} \\
T02 & 071 &  7.47 & 180 & 0.14 & 156 & 95 & 0.6 & 3.9 & 42 & 9 & \ding{52} & \ding{55} & \ding{52}\\ 
T03 & 012 & 7.12 & 260 & 0.25 & 163 & 88 & 0.7 & 2.8 & 58 & 24 & \ding{52} & \ding{55} & \ding{52}\\
T04 & 044 & 6.83 & 310 & 0.16 & 149 & 90 & 0.6 & 3.9 & 51 & 15 & \ding{52} & \ding{52} & \ding{52}\\
T05 & 001 & 6.85 & 360 & 0.21 & 149 &  85 & 1.2 & 5.4 & 37 & 37 & \ding{52} & \ding{52} & \ding{52}\\
T06 & 006 & 7.00 & 380 & 0.22 & 138 &  84 & 0.9 & 4.1 & 60 & 42 & \ding{55} & \ding{55} & \ding{52}\\
\hline 
\end{tabular}
\caption{Summary of calculated properties of \textit{representative} BC structures belonging to different families i.e. diamond (D), 
graphite (G), slab (S) and tubulane (T). The structures are represented with an id(first column), where the first
letter represents the family  and the last two integers, their energy ranking. The
space group indexes of the structures are listed in the second column. The quantities volume (\AA$^3$), energy($\Delta$E 
in meV) and electronic density of state at the Fermi level N($E_F$)(states/eV) are given per atom. For 
each family, the lowest-energy C structure and the  e $\alpha$-Rhombohedral-B$_{12}$ 
are considered as references for computing the formation energy of the structures. Quantities $\omega_{max}$ (maximum frequency at the
$\Gamma$-point), $\omega_{avg}$ (average of the optical vibrational frequencies at the $\Gamma$-point) calculated for a 8 atom unit cell 
and the logarithmic average phonon frequency $\omega_{log}$ are in meV. The \ep\; coupling constant $\lambda$ is dimensionless and the
\ep\; matrix element $\lambda$/N($E_F$) is in (states/eV/atom)$^{-1}$. The superconducting critical temperature \tc\; in K has
been estimated using the McMillan-Allen-Dynes formula \cite{PhysRev.167.331} with $\mu^*$= 0.10. 
The last three columns lists the presence (\ding{52}) or absence (\ding{55}) of C-C, B-B and B-C bonds respectively.}
\label{tab:ALLproperties}
\end{table*}

\section{Structures : Thermodynamics and Prototypes}
\label{sect:DB}
All structures considered in this work have a 50\%/50\% B/C stoichiometry, and can be
described with an 8-atoms unit cell (B$_{4}$C$_{4}$). This choice leaves out some
interesting structural prototypes, such as nanotubes and fullerenes, but is a reasonable
compromise between computational efficiency and structural flexibility.

Our initial MH runs produced around $\sim$ 630 such structures. Through post-relaxation 
of this initial pool with tighter settings and removal of duplicates, we ended up with 
a final tally of 320 unique structures. The Energy vs Volume plot of these structures is
shown in the upper panel of Fig.\ref{fig:volene}.
The energy shown here is the formation energy, computed using the graphite structure for
Carbon and $\alpha$-Rhombohedral-B$_{12}$ for Boron as references. All the BC structures
are metastable with positive formation energies in the range 0.1-1.0 eV/atom. Although large,
these values lie within the synthesizability threshold
defined in Ref.~\onlinecite{aykol2018thermodynamic} for carbon
polymorphs. Also, note that the formation energy may be sensibly reduced by choosing a different
initial carbon structure for the synthesis which is closer to the target candidate.

The colors and symbols in Fig.\ref{fig:volene} indicate  the different families  each
structure belongs to, i.e. blue triangles for graphite (G), inverted green triangles 
for slab (S), grey circles for diamond (D) and red circles for tubulane (T). The
classification of the structures into different families was done by hand, based on 
the relative arrangement of tetrahedral/triangular motifs and on the fraction of $sp^1$,
$sp^2$ and $sp^3$ bonds.

The combined variation of bonding fraction and the spatial arrangement of the tetrahedra
and triangles formed by $sp^3$ or $sp^2$ bonds make some of the BC structures different
from diamond or graphite. Their motifs are similar to those which have already been 
reported in pure carbon structures~\cite{lin2016interpenetrating,jiang2013tunable,          feng2018monoclinic,chen2015nanostructured,wang2018semimetallic,gao2018class, PhysRevMaterials.2.044205},
and classified as "interpenetrating graphene networks"
(\textit{IGN})\cite{lin2016interpenetrating} or
"carbon honeycombs" (\textit{CHC})\cite{PhysRevMaterials.2.044205}.
We have grouped these structures under the general keywords \textit{tubulanes}. 
In addition, we have created a new category \textit{slab}, to accomodate 
structures whose representative motifs have not been reported in literature till 
date.

The characteristic features of the four different families are summarized below: 
\begin{itemize}
    \item \textbf{Diamond} [ grey circles, Fig. \ref{fig:volene}(a) ] \\
    Diamond structures are characterized by dominant $sp^3$ bonding, which makes them
    occupy small volumes. Indicated by grey circles, the diamond structures are 
    situated on the left side in the Volume vs Energy plot in Fig.\ref{fig:volene}.
   
    \item \textbf{Graphite} [ blue triangle, Fig. \ref{fig:volene}(b) ] \\
    Graphite structures are characterized by atomically thin layers stacked on top
    of each other. The majority of atoms within one layer are bonded through $sp^2$ 
    bonding. The layers interact weakly through van-der Waals interaction. This makes
    them occupy large atomic volumes as indicated by the location of the blue triangles in the right half of Fig.~\ref{fig:volene}.
    
   \item \textbf{Slab} [ inverted green triangles, Fig. \ref{fig:volene}(c) ] \\
    A slab structure is geometrically similar to a graphite structure. However, at
    variance with graphite, formed by equispaced single layers, a slab structure is
    formed by \textit{slabs} of multiple atomic layers, separated by void. As shown
    in Fig.\ref{fig:volene}(c), each \textit{slab} comprises four atomic layers. 
    Despite having finite thickness, these slabs experience weak van-der-Waals
    interaction between them with an inter-slab distance of $\sim$2.9 \AA. The atoms 
    in this kind of system can form a mixture of $sp^2-sp^3$ bonds and hence occupy a
    large range of volumes. This is clearly evident from the large spread of the
    inverted green triangles as in Fig. \ref{fig:volene}.
    
    \item \textbf{Tubulane} [ red circles, Fig.\ref{fig:volene}(d) ] \\
    The word \textit{tubulane}, first reported by Baughman et. al. in
    1993, refers to structures which display 3D networks of tubular
    structures\cite{baughman1993tubulanes}.  These tubes can be of any shape i.e. rhombohedral,
    hexagonal, circular etc. The structures of the family \textit{IGN} and \textit{CHC}
    mentioned above fall in this category. As a typical example,
    Fig.\ref{fig:volene}(d) displays a tubulane with rhombohedral tubes. The constituent
    atoms in a tubulane can be connected via $sp^1$, $sp^2$, $sp^3$ bonds. With the
    possibility of having diverse mixture of bonds and tubes of different shapes,
    tubulanes can exhibit wide variability of atomic volumes, as shown by the wide distribution of the red circles in 
    Fig.\ref{fig:volene}.
\end{itemize}

Note that a proper estimate of the formation enthalpy
of BC structures should take into account not only
the B and C end members, but also intermediate compositions.
We thus evaluated the convex-hull of BC, including the 
icosahedral structure with B$_{13}$C$_2$ composition, which is the lowest phase on the extended hull according to Ref.~\onlinecite{jay2019theoretical}.
Taking the B$_{13}$C$_2$ phase into account, the formation 
enthalpies in Fig.~\ref{fig:volene} are shifted uniformly 40 meV up.
If, instead of the ground-state graphite-C, amorphous carbon
is considered as a reference, the BC composition falls back on the hull; the actual formation enthalpies are then 470 meV lower than in Fig.~\ref{fig:volene}
-- see Appendix B, Fig.~\ref{fig:Convexhull}. 
Many of the BC phases considered in this work may thus be synthesized, using the appropriate C precursor.

\section{Superconducting Trends of Representative Structures}
\label{sect:trends}
Superconductivity calculations are around two orders of magnitude more expensive than 
the total energy and structural relaxation runs used to construct our initial database 
of structures. In order to narrow down our pool of potential superconductors,
we first pruned out those structures, which have too high formation energies to be
synthesizable, are dynamically unstable or exhibit poor metallic character
and lack the stiff bonds, which are essential prerequisites for conventional HTS.

This was done by computing the value of electronic density of states (DOS) at the 
Fermi level N($E_F$) and the  vibrational frequencies at the zone center ($\omega_{i}$)
for all structure in the database(DB). These quantities, together with the formation 
energy $\Delta$E, were used to perform a preliminary screening, which left us with 116
potential candidates for HTS. As this number was still an order of magnitude
too large to afford full \tc\; calculations, we manually hand-picked sixteen 
representative candidates for accurate superconductivity calculations, shown as dark 
colored symbols in Fig.\ref{fig:volene}. In this second selection, we tried to choose
structures with diverse structural motifs and arrangements of B-C bonds. 
A detailed description of the screening protocol and the criteria of selection can be 
found in Appendix B.

The sixteen representative structures have been further relaxed with a 
Perdew-Wang-1992-LDA\cite{LDA-PW} functional before studying their geometric, electronic,
vibrational and superconducting properties. This second relaxation was needed, because it
is well known that structural and dynamical properties of layered (van-der-Waals systems)
systems are poorly described within GGA, while LDA gives a reasonable account; the explicit
inclusion of van-der-Waals corrections in DFPT calculations
of the \ep\; interaction is not yet implemented in any publicly available code.

\begin{figure}[!htbp]
\includegraphics[width=1.0\columnwidth,angle=0]{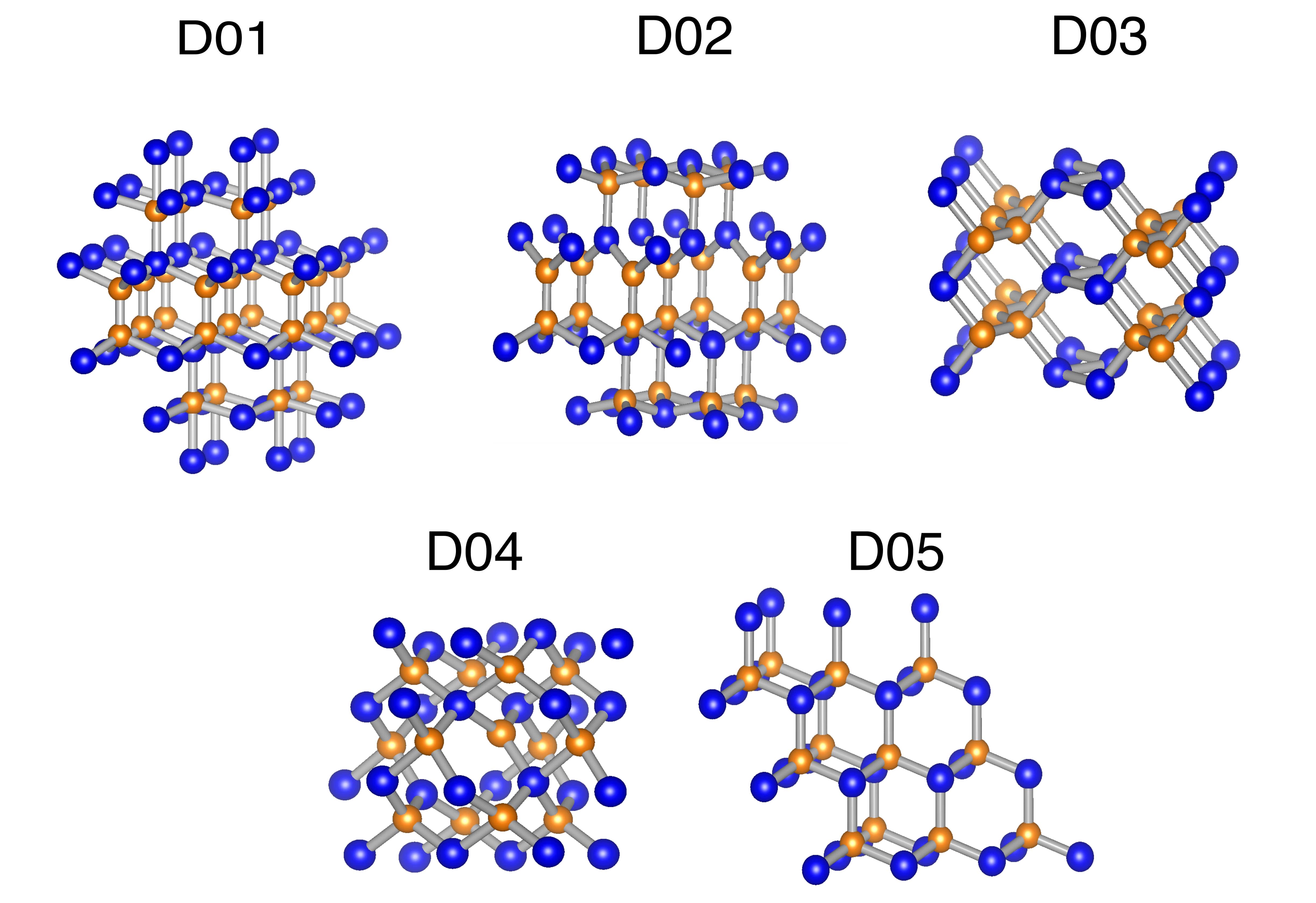}
\caption{{\em Representative} BC crystal structures in the diamond family. The B atoms 
are shown as  blue spheres, C atoms as orange spheres and bonds in grey. The structures 
are marked by their id.}
\label{fig:DDStruc}
\end{figure}

\begin{figure}[!htbp]
\includegraphics[width=1.0\columnwidth,angle=0]{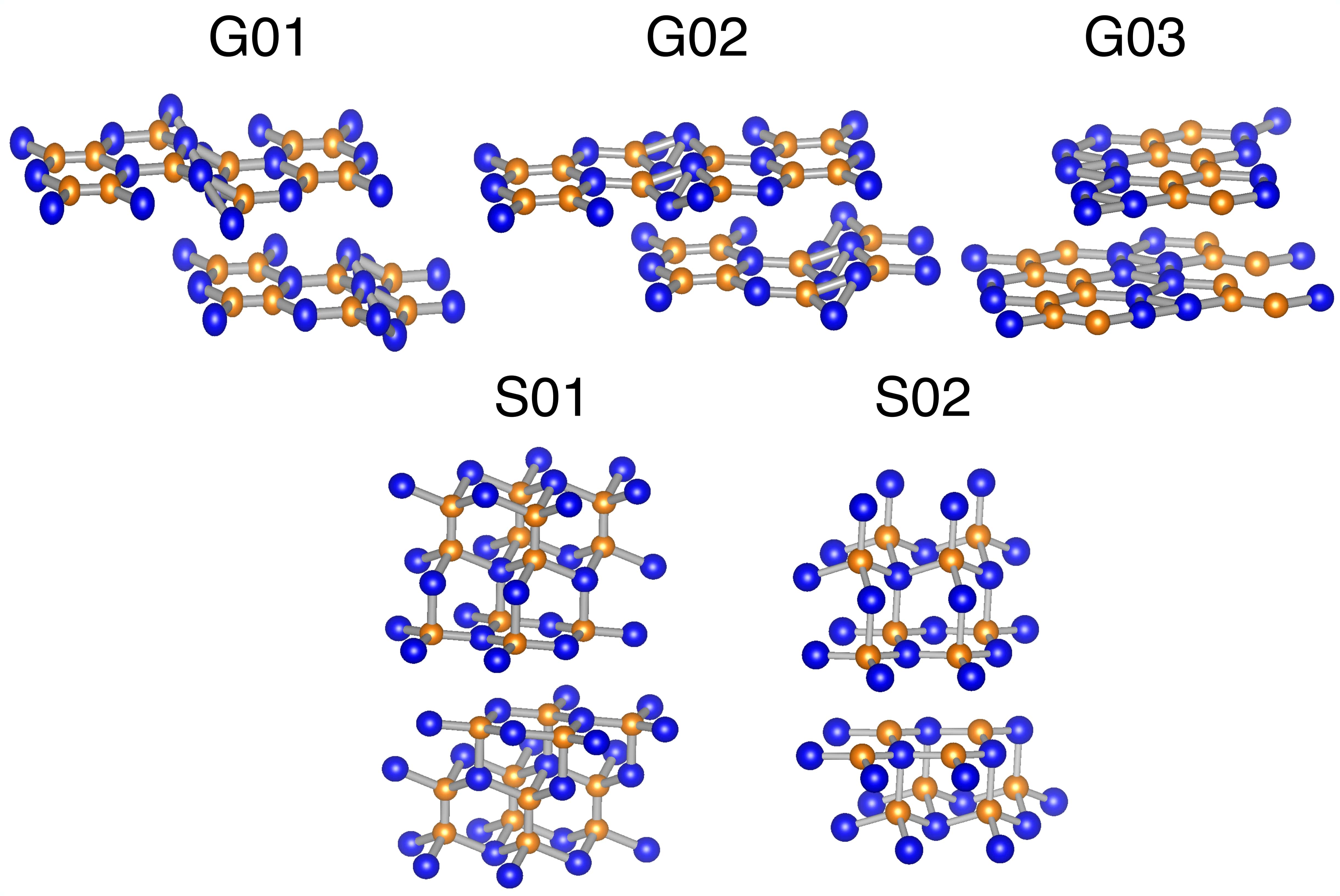}
\caption{{\em Representative} BC crystal structures in the graphite and slab family. 
The B atoms are shown as blue spheres, C atoms by orange spheres and bonds in grey. 
The structures are marked by their id.}
\label{fig:GGStruc}
\end{figure}

\begin{figure}[!htbp]
\includegraphics[width=1.0\columnwidth,angle=0]{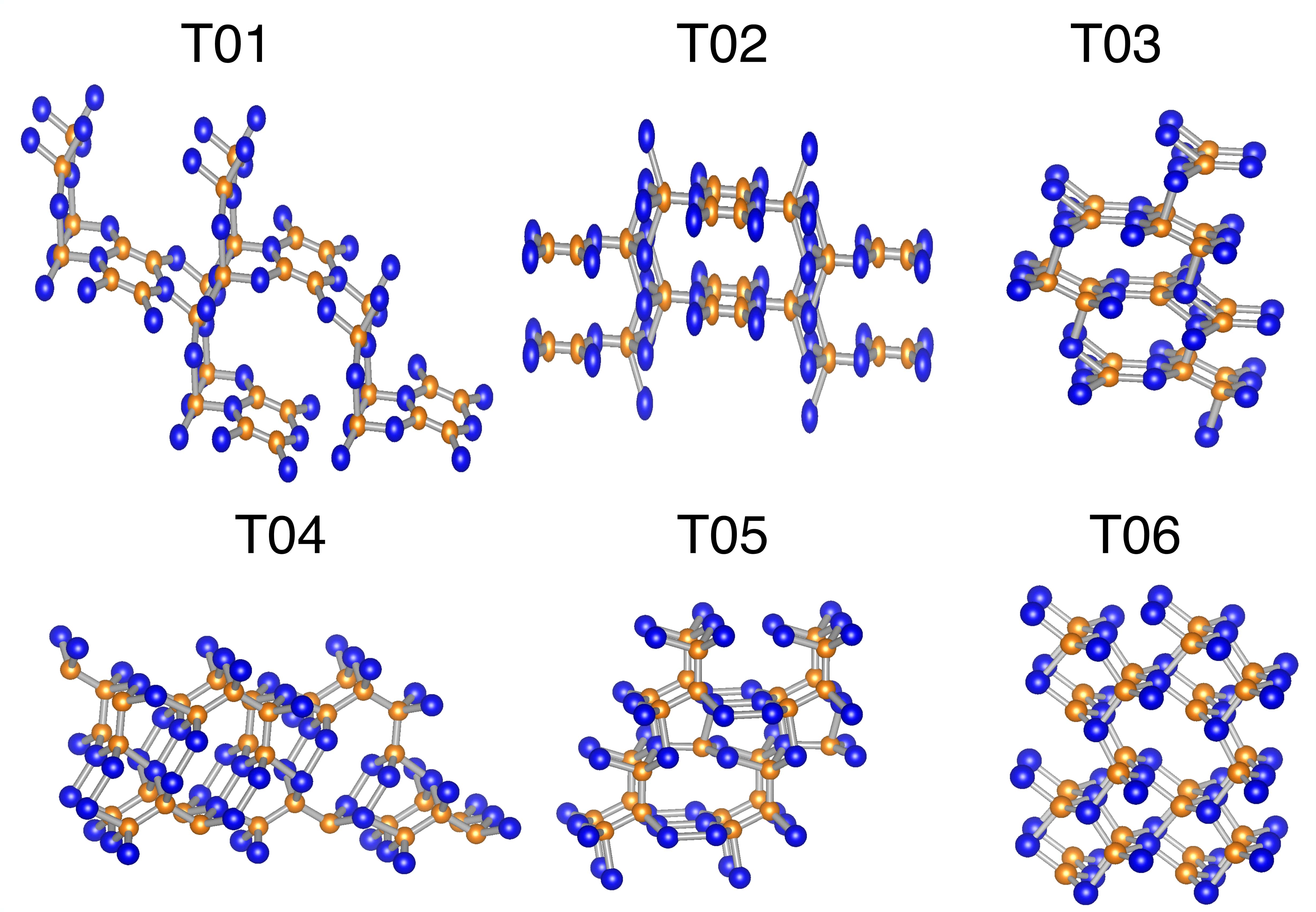}
\caption{{\em Representative} BC crystal structures in the tubulane family. The B atoms
are shown as  blue spheres, C atoms as orange spheres and bonds in grey. The structures 
are marked by their id.}
\label{fig:TTStruc}
\end{figure}

Grouped by family, the structures are shown in Figs.~\ref{fig:DDStruc}-
~\ref{fig:TTStruc} and their properties are summarized in Table
\ref{tab:ALLproperties}. In the following, the structures are represented by an
alphanumeric \textit{id} of the form A\textit{XY}, where the letter A represents the
family (Diamond, Graphite, Tubulane, Slab) and \textit{XY} the energetic ranking. The 
CIF files of the sixteen representative structures can be found in the supplementary 
material(SM).

The first column in Table \ref{tab:ALLproperties} lists the id of the selected structures.
General quantities describing the geometry are in the second 
(Space Group index), third 
(Volume per atom) and in the last three columns, which indicate the presence
(\ding{52}) or absence(\ding{55}) of bonds between C-C, B-B and B-C
respectively.~\footnote{The threshold of distance for the presence of C-C, B-B and B-C
bonds are d$_{CC}\le$ 1.5 \AA\;, d$_{BB}\le$ 1.85 \AA\; and d$_{BC}\le$ 1.65 \AA\;
respectively.} The formation energy ($\Delta E$) and the electronic DOS at the Fermi 
level N($E_F$) are in the fourth and fifth column respectively. Note that, in contrast to Fig.~\ref{fig:volene}, here the $\Delta E$
in each family is reported considering as reference
the lowest-energy C structure within that family. For the D and G families, 
diamond and graphite were considered. For the S family, 
diamond was considered. For the T family,
the structure \textit{IGN-Z33} from Ref.~\onlinecite{wang2018semimetallic} was considered,
which is a member of the \textit{IGN} family and the lowest energy (0.1 eV/atom w.r.t.
graphite) structure in the tubulane family. This structure has symmetric rhombohedral 
tubes where the 4 sides are made of 3 C chains arranged in zig-zag(ZZ) fashion, hence 
the name "Z33".

Column six and seven list the  ($\omega_{max}$) (meV) and the average ($\omega_{avg}$)(meV)
$\Gamma-$point vibrational frequency, evaluated on a 8 atoms unit-cell for all structures. 
The quantities $\Delta E$, N($E_F$) and $\omega_{avg}$ have been used for a pre-screening of
structures as discussed in Appendix B.

Quantities associated with superconducting properties listed in Table \ref{tab:ALLproperties}
are the \ep\; coupling constant $\lambda$, the approximate effective \ep\; matrix element 
$\lambda$/N($E_F$), the logarithmic average phonon frequency $\omega_{log}$(meV) and the
superconducting critical temperature \tc\;(K) estimated with the McMillan-Allen-Dynes formula\cite{PhysRev.167.331}:
\begin{equation}
    T_c=\frac{\omega_{log}}{1.2} exp \bigg[ - \frac{1.04(1+\lambda)}{\lambda-\mu^*(1+0.62\lambda)}\bigg],
    \label{eq:tcmcmillan}
\end{equation}
with a standard value $\mu^*$=0.10 for the Coulomb pseudopotential.
%%% CRYSTAL STRUCTURES 
%%%%%%%%%%%%%%%%%%%%%%%%%%%%%%%%%%%

\subsection{Structural Properties}
%%%%%%%%55DESCRIPTION OF DIAMOND STRUCTURES
The structures listed in Table~\ref{tab:ALLproperties} are shown in
Figs.~\ref{fig:DDStruc}-\ref{fig:TTStruc}.
All diamond structures, which consist of a mixture of $sp^2$ and $sp^3$ bonds, contain 
B-C bonds. Structures D01 and D02 also contain C-C bonds. The C-C bond is shared by two 
opposite-facing tetrahedra, while the other three bonds of the tetrahedron are B-C bonds.  
Structure D03 is the only structure in the D family which contains B-B bonds. The structure
consists of zig-zag chains of C and B ordered in a particular fashion to form $sp^3$ bonds.
Both D04 and D05 only involve B-C bonds. The major difference between the two is that
the structure D04 encompasses a mixture of $sp^2$ and $sp^3$ bonds, whereas D05
only contains $sp^3$ bonds. 

%%%%%%%%%%DESCRIPTION OF GRAPHITIC STRUCTURES
The layers of the graphitic structures G01 and G02 are the same, and the two structures
only differ in the relative arrangement of the layers. Unlike the layers of pure
C-graphite, which are flat, these layers have a staircase shape and consist of hexagons
formed by C-C  and B-C bonds. The B atoms which form the edge of the staircase have
coordination number $\sim$4-5, and hence form bonds which cannot be classified as purely
$sp^2$ or $sp^3$. The structure G03 contains flat atomic layers like graphite, in which
arm-chair chains of C atoms are connected to ordered chains of multi-bonded B atoms. These
two chains of C and B form hexagons and pentagons along with the clustering of B atoms.
Like every graphitic structure, there is a large inter-layer distance. 

%%%%%%%%%%%%DESCRIPTION OF SLAB STRUCTURES
The two slab structures S01 and S02 have identical structural templates;  
The only difference is that the B,C atoms in S02    
are arranged such that they only have
B-C bonds whereas S01 also has C-C bonds along with B-C bonds.
Each slab layer consists of 4 atomic layers; the two inner layers are bonded through 
$sp^3$ bonds whereas the 2 outer layers are bonded through $sp^3$ bonds with the inner layer,
while the remaining bonds are $sp^2$-like. The two outer layers contain hexagons.

%%%%%%%%%%%DESCRIPTION OF TUBULANE STRUCTURES
The \textit{representative} structures of the tubulane family all encompass 3D tubes of
different shapes and sizes, with different fractions of $sp^2$-$sp^3$ bonds. Tubes with
large diameter occupy larger volumes, as seen in T01 and T02. Like every other structural
templates, also in the tubulane family all  members contain B-C bonds; in addition, all
members except T06 contain C-C bonds. The C atoms in C-C bonds are part of an $sp^2$
geometry in structure T01 and T02. In the remaining cases, they are in an $sp^3$ geometry.
Structures T04 and T05 contain B-B bonds which are part of buckled hexagons, arranged in 
a $sp^2$-$sp^3$ geometry. Structure T06 only contains B-C bonds, which are in a $sp^3$
geometry.

\subsection{Trends in \tc\;}
The sixteen structures listed in Table~\ref{tab:ALLproperties} represent a diverse 
sample of possible structural motifs and properties. Before analyzing their electronic
structure in detail, some general trends across and within families can already
be discussed on the basis of the data in table~\ref{tab:ALLproperties}.

In general, three observations are in place:
\begin{enumerate}
\item
In all families except graphite, we found structures with rather high values of the 
DOS at the Fermi level N($E_F$), and moderate to high-\tc's. We also observe that 
these structures with higher DOS and \tc\; tend to have quite high formation energies
$\Delta E$, of the order of 200-400 meV, which is close to the synthetizability threshold.
\cite{aykol2018thermodynamic}
%({\bf Santanu, is this a general trend of the whole DB?}NO, THE STRUCTURES WITH LARGE DOS
%ARE IN THE INTERMEDIATE RANGE OF ENERGY 0.1-.0.6 eV/atom).
The structure with the highest N($E_F$) is D05, which is a diamond structure with only B-C
bonds in a perfect tetrahedral geometry. Other structures with high N($E_F$) are D04, S02, 
T05 and T06, which all exhibit \tc's exceeding 30 K. All graphitic structures obtained from 
MH runs have rather small values of N($E_F$), and negligible \tc's. We thus tried to 
construct graphitic structures with high N($E_F$) manually, through different homogeneous
replacements of B in C graphite in a 8-atoms cell. However, we found that any arrangement 
of B atoms in C graphite induces buckling, and that these buckled structures are either
dynamically stable non-metals or dynamically unstable metals. This observation confirms what has
been observed in studies of B/N doping of single graphene sheets by Zhou et. al. and 
Mann et. al. ~\cite{PhysRevB.92.064505,mann2016thermodynamic}

\item A second quantity exhibiting a remarkable correlation with the \tc\; is
the value of the highest vibrational frequency at the $\Gamma$-point ($\omega_{max}$)
and, in particular, its reduction (softening) with respect to the same quantity in 
a reference structure of pure carbon. In general, the softening is more pronounced 
for structures with higher N($E_F$) and $\lambda$.
Almost all diamond structures exhibit a remarkable softening of the highest 
vibrational frequencies ($\omega_{max}$), with respect to that of pure diamond (164
meV).\footnote{The vibrational frequencies are calculated for the pure diamond
structure with the lattice constant of D05 i.e. a=3.74 \AA\;, which is a $\sim 5\%$
larger than the experimental lattice constant of diamond.}. The softening is the
highest for D05, where $\omega_{max}$ is reduced by a factor 0.65 compared to the
reference value. On the contrary, graphite structures exhibit only a small softening,
as compared to the the reference value for C graphite (195 meV). Tubulane structures
also exhibit a strong softening of $\omega_{max}$, compared to the reference tubulane
structure (ING-Z33 200 meV). It is hard to give a quantitative estimate of this effect
for slab structures, because no dynamically-stable reference structure exists, but 
the reference value should lie somewhere between $sp^3$ diamond and $sp^2$ graphite,
and both S01 and S02 exhibit a remarkable softening with respect to this value.

\item A third, more general correlation can be found across the whole database 
between \tc\; and the types of bonds (B-B, B-C or C-C) present. 
In particular, structures which contain B-C bonds only have the highest \tc\; 
within each family. Structures with/without C-C and B-B bonds along-with B-C bonds 
may or may not be superconductors. A close look at
Figs.~\ref{fig:DDStruc}-~\ref{fig:TTStruc} show that in structures containing B-B bonds,
the B atoms  are not part of $sp^2-sp^3$ bonds, but form
multiple bonds. This leads to a sizable deformation of the structure, which reduces the symmetry, and causes a sensible reduction of  $N(E_F)$, and hence \tc.
For example, B atoms in G01, G02 and G03 form 4-5 bonds with
both B and C. On the other hand, the role of C-C bonds in determining \tc\;
is much less clear. Finally, it is interesting to note that, though both D04 and D05
only contain B-C bonds, they have different \tc\;'s. This difference can be associated
to the fact that D05 only contains $sp^3$ bonds, whereas D04 contains a mixture of
$sp^2$-$sp^3$ bonds, and also in this case the symmetry lowering leads to a \tc\;
suppression.
\end{enumerate}

\section{Electronic Structure:}
\label{sect:elestruct}
In this section, we present a detailed comparison of the electronic structure of
the {\em representative structures}, to gain a microscopic insight of their
superconducting properties, discussed only in general terms so far.

The electronic DOS's, Phonon and electron-phonon spectra (Eliashberg functions) 
for our sixteen \textit{representative} structures, divided by families, are 
reported in Appendix B, Figs.~\ref{fig:DDele}-\ref{fig:TTa2F}; in the electronic
(Figs.~\ref{fig:DDele}-\ref{fig:TTele}) and phonon
(Figs.~\ref{fig:DDph}-\ref{fig:TTph}) DOS plots, we report in red and blue the 
partial carbon and boron contributions as well as the total DOS in black. 
The top panels of all figures show reference spectra, calculated for a pure carbon
structure.

The electronic DOS plots show an almost perfect hybridization between B and C states
in all structures, with the two partial DOS's closely following each other. 
In addition, the variation of the spectral distribution of the electronic
states in different BC structures, compared to the reference pure
carbon structures, is a good indicator of the changes in electronic structure due to
rearrangement of bonds.
In this respect, it is quite interesting to compare the behavior of structures in 
the diamond and graphite families, where it is straightforward to define a reference
template for the pure structure.
In both cases, in a simple rigid-band (RB) model the Fermi level, shown  by the 
dashed line in the upper panels of Figs.~\ref{fig:DDele}-~\ref{fig:GGele}, would 
fall into a $\sigma$ (2D or 3D) band. In this case, one would predict a sizable 
\ep\; coupling, as $\sigma$ bonds are extremely stiff and sensitive to lattice
distortions.~\cite{SC:diamond_Boeri_PRL2004}

However, in most real structures, a substantial rearrangement of bonds and electronic states invalidates this simple line of reasoning, based 
on the the RB approximation. 
In the diamond family, a substantial shift of spectral weight away from the Fermi 
level occurs, which is more pronounced for low-energy structures, where it produces
a substantial lowering of the DOS at the Fermi; the shift is absent in D05, which
can almost perfectly be described by the RB approximation.
In the graphite family, all structures generated for 50$\%$ BC composition 
are either dynamically unstable, or weakly metallic, due to a major rearrangement 
of bonds. G01, G02 and G03 all contain B-B and C-C dimers, and/or buckled planes, 
and exhibit an extremely small $N(E_F)$.

For slabs and tubulanes, due to the large variety of moieties and motifs,
it is less straightforward to define a reference structure. We
chose T06 and S02 as structural template for the C reference structure for
tubulane and slab respectively. Also in these cases, a pronounced shift of 
spectral weight away from the Fermi level is observed, which is reduced for
higher-energy structures. The DOS of low-energy tubulanes, which are more open, 
resemble quite closely those of graphite structures, while high-energy ones
tend to mimic those of diamond. The same tendency can be observed in slab 
structures.

Phonon DOS's are shown in Figs.~\ref{fig:DDph}-\ref{fig:TTph}, again
with the same color-code and definition of reference structures.
As observed for electronic DOS's, due to the similar B and C mass, the spectra have
in general a fairly mixed character. However, the phonon DOS's of structures which
contain B-B or C-C bonds tend to exhibit sharp  peaks of pure B- or C- character,
corresponding to localized vibrations. Many of these peaks are found at high 
energies.
In addition, a progressive reduction of the highest phonon frequency with increasing
formation energy is also evident in all families. The effect is particularly
spectacular in DO5, where the reduction of the highest frequency is $\sim 35 \%$.

Figs.~\ref{fig:DDa2F}-\ref{fig:TTa2F} show for each family the Migdal-Eliashberg(ME)
\ep\; spectral function:\cite{eliashberg1960interactions}
\begin{equation}
    \alpha^2F(\omega)= \frac{1}{N(E_F)} \sum_{\textbf{kq},\nu} |g_{\textbf{k},\textbf{k+q},\nu}|^2 \delta(\epsilon_\textbf{k}) \delta(\epsilon_\textbf{k+q})
    \delta(\omega-\omega_{\textbf{q},\nu})
\label{eq:alpha}
\end{equation}
where N($E_F$) is the electronic DOS at the Fermi level, and the two $\delta$ 
functions restrict the sum to electronic states at the Fermi level with momenta \textbf{k} and
\textbf{k+q}. The $\omega_{\textbf{q},\nu}$ is the vibrational
frequency of mode $\nu$ and wavevector $\textbf{q}$ and
$g_{\textbf{k},\textbf{k+q},\nu}$ is the corresponding electron-phonon matrix element.
On the same plots, with orange dashed-lines we show the frequency-depent \ep\; 
constant $\lambda$($\omega$) and report the average phonon frequency $\omega_{log}$,
given by:
\begin{equation}
    \lambda(\omega)=2 \int_0^\omega   \frac{\alpha^2F(\omega')}{\omega'}d\omega'
\label{eq:lambda}
\end{equation}
\begin{equation}
    \omega_{log}= exp \bigg[ \frac{2}{\lambda}\int_0^\infty \alpha^2F(\omega)\frac{ln (\omega)}{\omega}d\omega \bigg]
\label{eq:omegalog}
\end{equation}
which measure respectively the average energy of the phonons which couple mostly 
to electrons, and of the intensity of the \ep\; coupling.

In most compounds, the Eliashberg function is almost proportional to the phonon
DOS, reflecting a uniform spread of the \ep\; coupling on the phonon spectrum. A
notable exception is the slab structure S02, where there is a substantial 
enhancement of coupling to phonons in the low-energy region. While the values of
$\omega_{log}$ are quite spread out, without any clear trend for low- or high-energy
structures, the values of the total \ep\; coupling constant $\lambda$, obtained from
Eq.~\ref{eq:lambda} with $\omega=\infty$, tend to be larger for higher-energy
structures, and range from 0.4 in graphite structures G01, G02 and G03 to 2.3 in 
diamond D05. 

The main factor behind the large variation in $\lambda$ amongst structures is the
variation of the electronic DOS at the Fermi level N$(E_F)$. This can be appreciated
recalling that $\lambda$  can be rewritten using the so-called Hopfield
expression:\cite{PhysRev.186.443}
\begin{equation}
    \lambda = \frac{N(E_F) I^2}{M \tilde{\omega}^2},
\label{eq:Hopfield}
\end{equation}
where $I^2$ is the \ep\; coupling matrix element averaged over Fermi surface, M is 
the average atomic mass and $\tilde{\omega}^2$ is the square of an average 
vibrational frequency. 
As reported in Table ~\ref{tab:ALLproperties}, $V=\frac{\lambda}{N(E_F)}$ is $\simeq 4.0$ in most structures considered in this work.
The only notable exception is the diamond structure D05, where this ratio is 
50 $\%$ larger than in all other structures, reflecting a qualitative difference
in bonding with respect to all other structures.

In summary, the analysis of the electronic structure shows that most structural
templates exhibit a similar tendency to superconductivity: the \ep\; coupling is
spread out over several phonon modes, and the value of the \ep\; coupling constant
$\lambda$, and hence \tc, is mostly determined by the value of N($E_F$), since the
ratio $V=\lambda/N(E_F)$, is essentially constant across and within families. In 
most low-energy structures \tc\; is suppressed by the formation of B-B and C-C bonds,
which shifts electronic spectral weight away from the Fermi level, lowering the band
energy, but also N($E_F$).

The presence of B-B and C-C bonds is also visible in the phonon spectra, where it leads
to the formation of sharp peaks at high energies.

The diamond structure D05, where, due to the alternating arrangement of B-C atoms, the
original symmetry of pure diamond is retained, and electronic states at the Fermi level
have a pure $\sigma$ ($sp^3$) character, is a clear outlier of the database. Here, the
DOS follows a perfect rigid-band behavior compared to pure diamond, while the phonon
spectrum is strongly renormalized, due to coupling between bond-stretching phonons and
$\sigma$ states. As a result, $V=\lambda/N(E_F)$ is around 50 $\%$ larger than in all
other representative structures, and %
the predicted superconducting \tc\; is also exceptional ($79~K$), in line with the
highest values calculated in
Ref.~\cite{Moussa_diamond_PRB2008}.

These observations imply that general arguments 
based on the rigid-band analysis of fixed structural
templates must be taken with care in BC,~\cite{SC:Moussa_molfrag_PRB2008}
because structural
distortions and bond rearrangements can have a dramatic effect on \tc.

\section{A simple Expression for \tc:}
\label{sect:Tcformula}
The values of \tc, $\lambda$ and $\omega_{log}$ for all sixteen representative
structures, collected in Table~\ref{tab:ALLproperties}, were computed using the
McMillan-Allen-Dynes formula, Eq.~\ref{eq:tcmcmillan}, which requires a full
calculation of the electron-phonon (Eliashberg) spectral function - 
Eq.~\ref{eq:alpha}.

For an 8-atoms unit cell with no symmetry, a calculation of $\alpha^2 F(\omega)$ 
with a reasonably-dense sampling of  reciprocal space for electronic and phononic
momenta requires around 3000 CPU hours on a computer cluster. This type of calculations are 
clearly unfeasible for large-scale high-throughput material screening, which was our 
primary motivation to pre-select only a few \textit{representative} structures from
our initial pool.

%%%%%%%FIG 2 - Tc vs N(Ef)/W_max PLOT
\begin{figure}[!thb]
\includegraphics[width=1.0\columnwidth,angle=0]{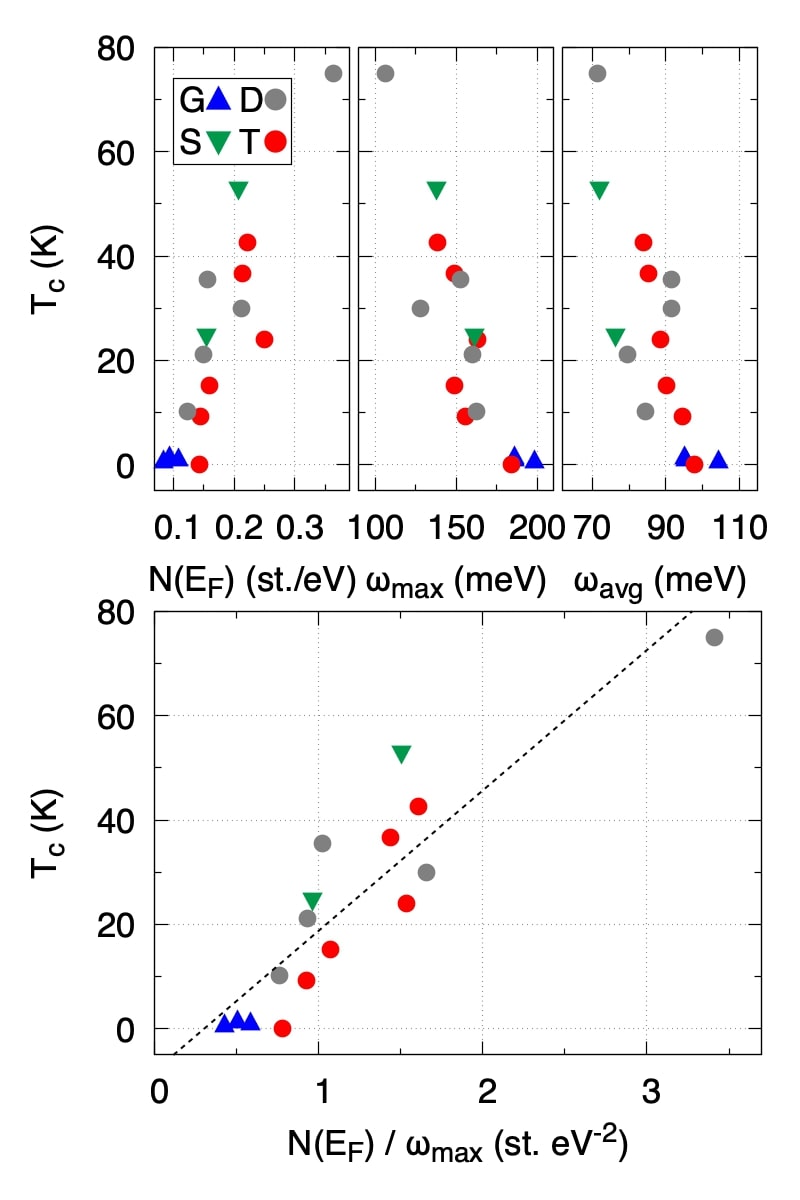}
\caption{(Top panel): Superconducting critical temperature \tc\; (K) of 
the sixteen {\em representative} BC structures as a function of: (left)
electronic DOS at the Fermi level
N($E_F$)(states/eV/atom) left, (middle) maximum vibrational
frequency at the $\Gamma$-point $\omega_{max}$(meV)); (right) average optical vibrational frequency at the $\Gamma$-point
$\omega_{max}$(meV)).
(Bottom panel): The \tc\;'s of the sixteen {\em representative} BC structures are plotted
as a function of N($E_F$)/$\omega_{max}$ in (states/eV$^2$/atom). The colour and the  
symbols in the plot indicate the family each structure belongs to. 
The dotted line represents an approximate linear fit to  the \tc; data - Eq.~\ref{eq:tcline}.}
\label{fig:TcNefWmax}
\end{figure}

With the \tc\; data at hand, it is  interesting to see whether any trends in \tc\;
could have been foreseen on the basis of the simple electronic structure quantities
that we had used to pre-screen our structural database, which require a much less
intense computational effort.

The three upper panels of Fig.~\ref{fig:TcNefWmax} show that \tc\; exhibits an 
almost linear correlation with N($E_F$), and an inverse correlation with both $\omega_{max}$ and $\omega_{avg}$.
Although the two vibrational descriptors are approximately equivalent, $\omega_{max}$ is monotonous, while
 $\omega_{avg}$ incorrectly classifies the two slab
 structures and a few diamond ones.
The lower panel of Fig.~\ref{fig:TcNefWmax} shows that the calculated \tc's when plotted as a 
function of $\frac{N(E_F)}{\omega_{max}}$ closely follow a linear behavior:

%a=26.8702
%b=-8.03291
%f(x)=a*x+b

\begin{equation}
T_c= 26.9 K st.^{-1} eV^2\cdot\left[\frac{N(E_F)}{\omega_{max}}-0.3\right] 
\label{eq:tcline}
\end{equation}
 
Although extremely simple, this formula seems to interpolate nicely the \tc\; 
from different templates, and has a transparent physical interpretation.

That \tc\; should positively correlate with N($E_F$) can be easily understood from the Hopfield's
expression for $\lambda$ - Eq.~\ref{eq:Hopfield}. On the other hand, the correlation
of \tc\; with $\omega_{max}$ is less straightforward to understand.
One aspect is probably phonon softening: in an interacting system of phonons and
electrons, the same coupling which leads to superconductivity also leads to the
renormalization of phonon frequencies, with respect to a bare, non-interacting value.
In a simple model where a single phonon mode with frequency $\omega$ couples to a
single electronic band, $\omega$ is reduced with respect to its bare value 
$\Omega$ as: $\omega^2=\Omega^2(1-2\lambda)$.
However, while the model of a single phonon mode may be safely applied to hole-doped
diamond and graphite, where superconductivity is dominated by the zone-center
bond-stretching optical phonon and $\sigma$ holes,~\cite{SC:diamond_Boeri_PRL2004}
its applicability to structures where the RB model does not hold due to major
structural rearrangements, and  the coupling is spread out over several
phonon modes and electronic states, is questionable.
It is in fact possible that $\omega_{max}$ accidentally encodes both the presence 
of high-energy B-B or C-C phonon modes, due to the formation of B-B and C-C bonds,
which are disruptive for HTCS, and the actual phonon softening of a large part of
the phonon spectrum in systems where the coupling is strong. More tests are needed
to check the general validity of this trend, even for a relatively specialized case
as BC. This goes well beyond the aim of the present paper.
\begin{figure}[!thb]
\includegraphics[width=1.0\columnwidth,angle=0]{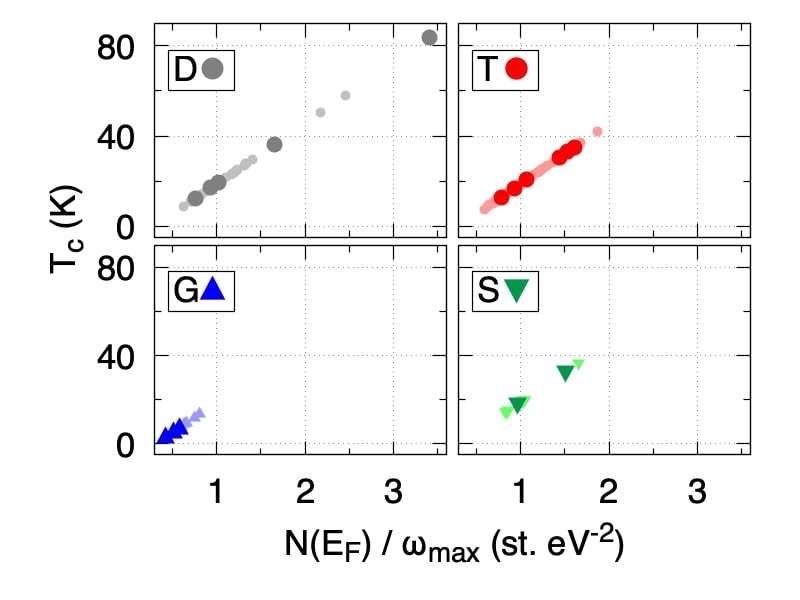}
\caption{The four panels show the model \tc (eq.\ref{eq:tcline}) w.r.t
N($E_F$)/$\omega_{max}$ in (states/eV$^2$/atom) for all the structures separated
by families. Here, N($E_F$) is the DOS at the Fermi level and $\omega_{max}$ is
the maximum vibrational frequency at the $\Gamma$-point. The top left panel is for
diamond (grey circles), top right for tubulane (red circles), bottom left for
graphite(blue triangles) and bottom right for slab(inverted green triangles). The
selected structures are shown by big dark coloured symbols. The rest are shown by small light coloured symbols.}
\label{fig:Tcalldata}
\end{figure}

However, we can use our simple predictor for \tc\;  to estimate the tendency of BC structures to HTCS across the
whole energy landscape. 
In Fig.\ref{fig:Tcalldata}, the model \tc\; 
from Eq.~\ref{eq:tcline},
as a function of N($E_F$)/$\omega_{max}$
is shown in four different panels for all metallic structures in the original DB, grouped by family.
Around 60 $\%$ of the predicted \tc\; lie in the 10-20 $K$
range, 25 $\%$ between 20 and 30 $K$, and $\sim$10$\%$ are above 30 $K$. These high-\tc\; structures belong 
mostly to the tubulane
and diamond families, a few to the slab family,
whereas all graphite structures  
 are predicted to exhibit \tc's below 20 $K$.
Note that also in this plot the structure
D05, i.e.  the data point with the highest \tc\; in Fig.\ref{fig:Tcalldata}, is a 
complete outlier, and most likely its \tc\; of 79 $K$ is an upper bound
for the BC system at 50$\%$-50$\%$ composition.

\begin{figure}[!thb]
\includegraphics[width=1.0\columnwidth,angle=0]{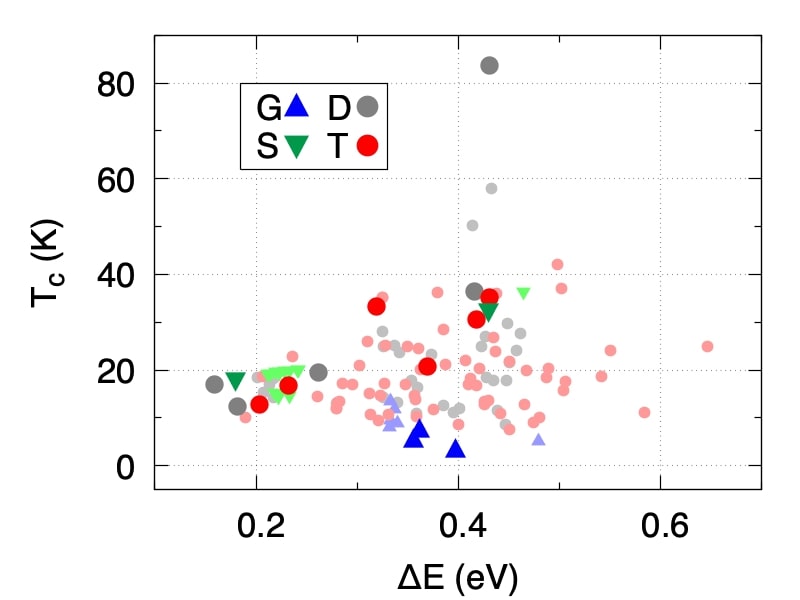}
\caption{Superconducting critical temperature \tc\; (K) of 
all the metallic BC structures as a function of formation
energy in (eV/atom). The colour and the symbols in the plot indicate the 
family each structure belongs to.}
\label{fig:Tcalldata-dE}
\end{figure}

Eq.~\ref{eq:tcline} can also be used to obtain an estimate of the formation energy
required to obtain structures for a specific range of \tc\;. In Fig.\ref{fig:Tcalldata-dE}, 
the predicted \tc\; for all metallic BC structures is plotted as a function of formation 
energy; the meaning of colors and symbols is the same as in Fig.~\ref{fig:volene}.
We observe that, in general, the lowest-energy structures have \tc's below 20~K. 
A large cluster of structures is also found, with $\Delta E$ between 300 and 400 meV,
and \tc's exceeding 30~K.

\section{Conclusions:}
\label{sect:conclusions}

In this work, we probed the possibility of realizing HTCS in the boron-carbon system, using an {\em ab-initio} screening approach. First, we generated a large 
(320) database of metastable BC structures, with 50$\%$/50$\%$ boron/carbon 
composition and 8-atoms unit cells, and showed that these can be grouped into
four main families of characteristic motifs for pure carbon: diamond, graphite, 
slab and tubulane. 
From a first high-throughput screening based on the values 
of the electronic DOS at the Fermi level, zone-center vibrational frequencies, 
and formation energies, we estimated that around half of the generated structures 
are promising HCTS. From these, we selected sixteen {\em representative} structures,
spanning a variety of motifs and structural templates, for which we performed full
electron-phonon calculations. We identified several general trends amongst them: 
($i$) In all families, except graphite, we could find superconductors
with \tc's $\simeq 40~K$, comparable to the best-known ambient-pressure
superconductors; ($ii$) Within one family, the value of \tc\; is essentially 
determined by N($E_F$); ($iii$) \tc\; correlates inversely with the highest phonon
frequecty at the zone-center, $\omega_{max}$. 
($iv$) A geometric analysis of the selected
structures shows that the highest \tc's within
a given family is usually found in structures
where the fraction of B-C bonds is dominant with respect to other types of bonds, and particularly if these have $sp^3$ character. Structures where bonds centered
around B atoms are neither $sp^2$ nor $sp^3$ tend to exhibit a low \tc\;, because the
clustering of atoms around B tends to reduce the symmetry, depress the value of N($E_F$), and hence \tc\;.

The empirical observations ($i$)-($iii$) can be distilled into a single analytical formula for \tc, which can be used as a predictor for HTCS. On the basis of this formula, we
estimate that around $\sim$10$\%$ structures have \tc\;
larger than 30~K, which makes them interesting candidates for HTCS. Most of these structures
have formation energies between 300 and 400 meV, and may be synthesized using an appropriate
carbon precursor. The diamond structure D05, which only has B-C bonds in $sp^3$ tetrahedral
arrangement, sets an upper limit for \tc $\sim$ 80 $K$ for the BC system at  50$\%$-50$\%$
composition. Given that \tc\; is so strongly dominated by
$N(E_F)$, it is however conceivable that \tc\; may be improved by doping.  
This work is a first step in the identification of HTCS at ambient pressure in
light-element covalent metals using {\em ab-initio} screening techniques.

\begin{acknowledgements}
S.Saha, S. di Cataldo and W. von der Linden acknowledge computational resources from
the dCluster of the Graz University of Technology and the VSC3 of the Vienna 
University of Technology, and support through the FWF, Austrian Science Fund, Project
P 30269- N36 (Superhydra). M. Amsler acknowledges support from the Swiss National Science
Foundation (projects P300P2-158407, P300P2- 174475, and P4P4P2-180669).
L. Boeri acknowledges support from Fondo Ateneo Sapienza
2017-18 and computational Resources from CINECA, proj. Hi-TSEPH.
\end{acknowledgements}

\clearpage

\appendix 

\section{Electron-Phonon Spectra of the Sixteen Representative structures}

\subsection{Electronic Properties}
%%%%%%%%%%%%%%%%%%%%%%%%%%ELECTRONIC PROPERTIES %%%%%%%%%%%%%%%%%%%%%%%%%%%%%
\begin{figure}[!htbp]
\includegraphics[width=0.9\columnwidth,angle=0]{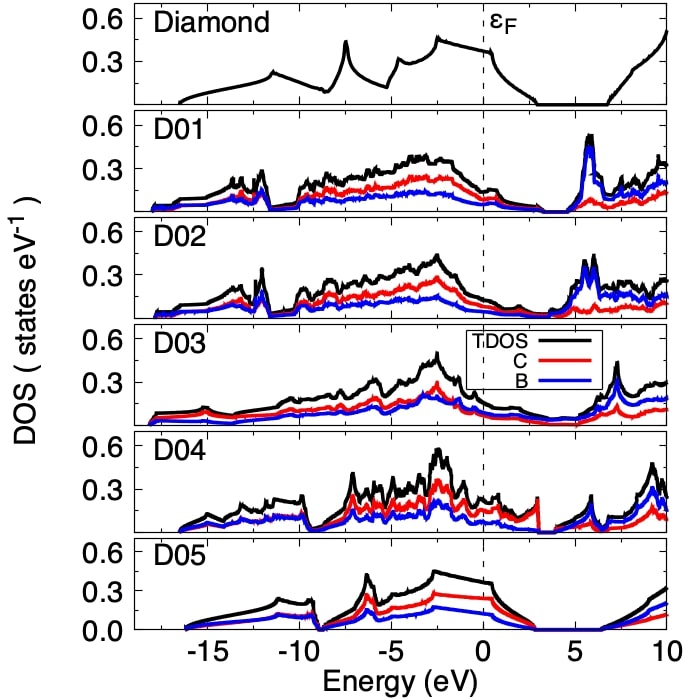}
\caption{The top panel shows the electronic DOS of the reference diamond structure and rest,
the total DOS(black) and partial contribution of C(red) and B(blue) atoms in all the diamond
structures.}
\label{fig:DDele}
\end{figure}

\begin{figure}[!htbp]
\includegraphics[width=0.9\columnwidth,angle=0]{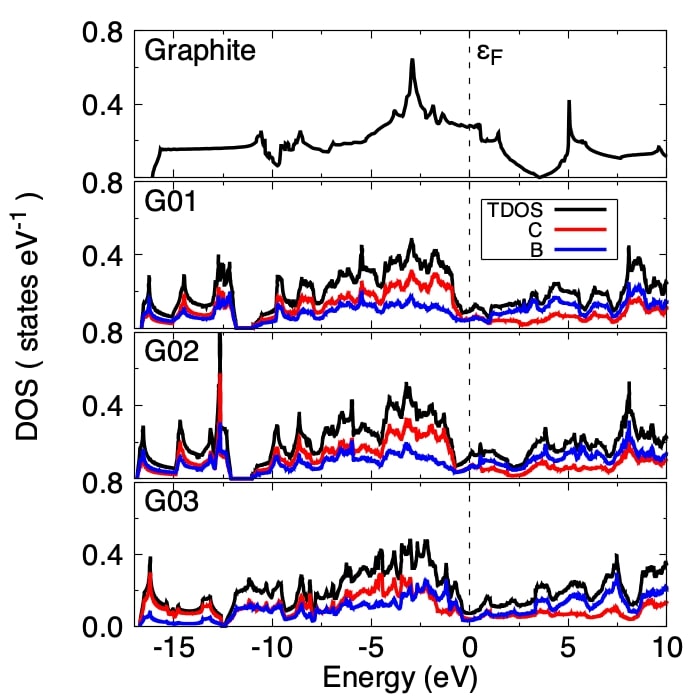}
\caption{The top panel shows the electronic DOS of the reference graphite structure and rest,
the total DOS(black) and partial contribution of C(red) and B(blue) atoms in all the graphite
structures.}
\label{fig:GGele}
\end{figure}

\begin{figure}[!htbp]
\includegraphics[width=0.9\columnwidth,angle=0]{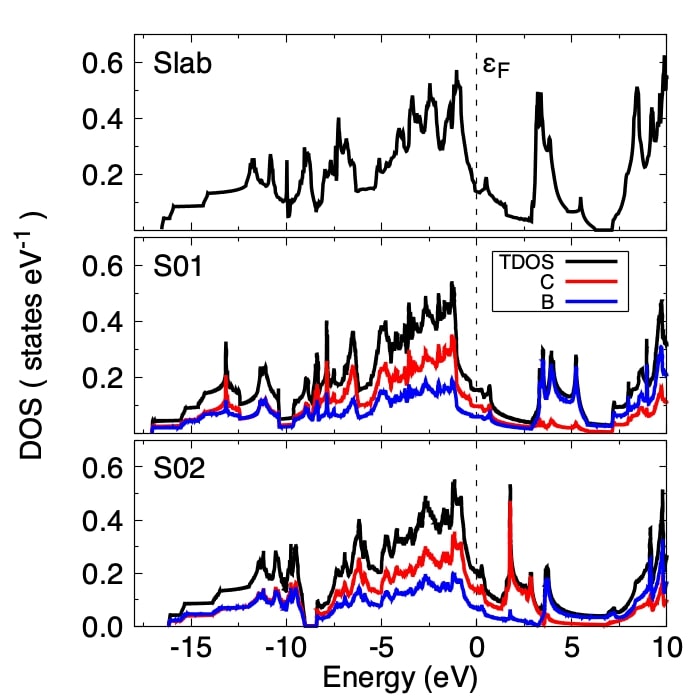}
\caption{The top panel shows the electronic DOS of the reference slab structure and rest,
the total DOS(black) and partial contribution of C(red) and B(blue) atoms in all the slab
structures.}
\label{fig:SSele}
\end{figure}

\begin{figure}[!htbp]
\includegraphics[width=0.9\columnwidth,angle=0]{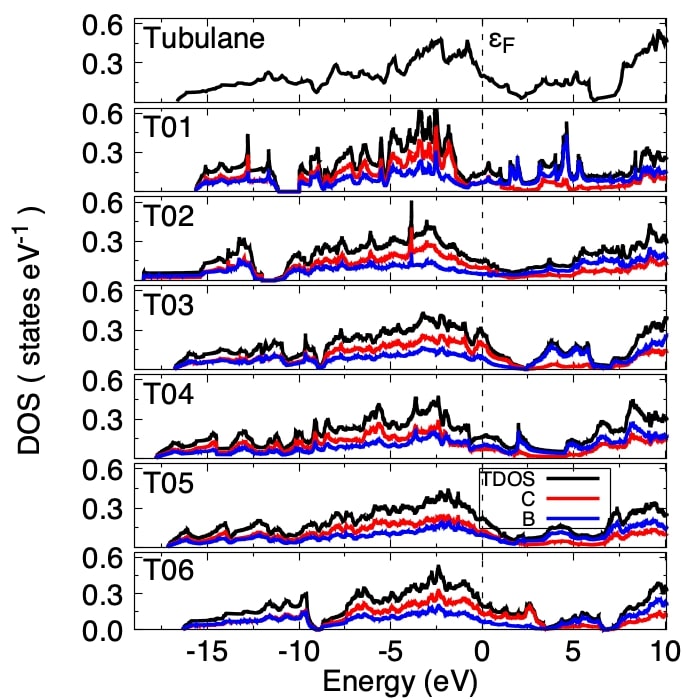}
\caption{The top panel shows the electronic DOS of the reference tubulane structure and rest,
the total DOS(black) and partial contribution of C(red) and B(blue) atoms in all the tubulane
structures.}
\label{fig:TTele}
\end{figure}

%\clearpage 
%%%%%%%%%%%%%%% VIBRATIONAL PROPERTIES
\subsection{Vibrational Properties}
\begin{figure}[!htbp]
\includegraphics[width=0.9\columnwidth,angle=0]{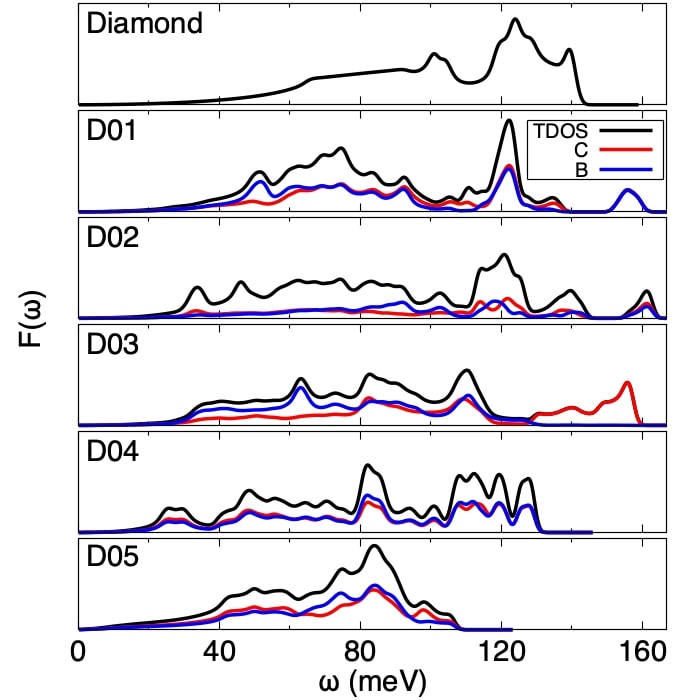}
\caption{The top panel shows the phonon DOS F($\omega$) of the reference diamond structure and
rest, the total DOS(black) and partial contribution of C(red) and B(blue) atoms in all the 
diamond structures.}
\label{fig:DDph}
\end{figure}

\begin{figure}[!htbp]
\includegraphics[width=0.9\columnwidth,angle=0]{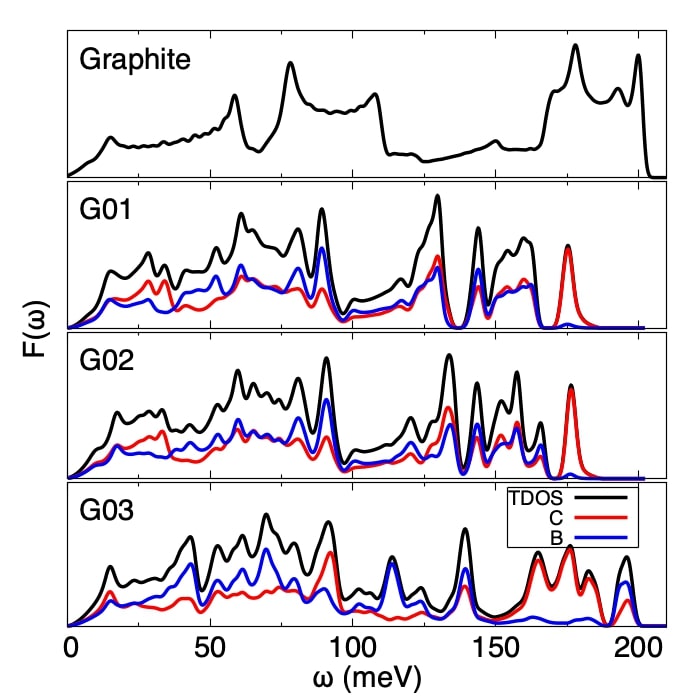}
\caption{The top panel shows the phonon DOS F($\omega$) of the reference graphite structure and
rest, the total DOS(black) and partial contribution of C(red) and B(blue) atoms in all the
graphite structures.}
\label{fig:GGph}
\end{figure}

\begin{figure}[!htbp]
\includegraphics[width=0.9\columnwidth,angle=0]{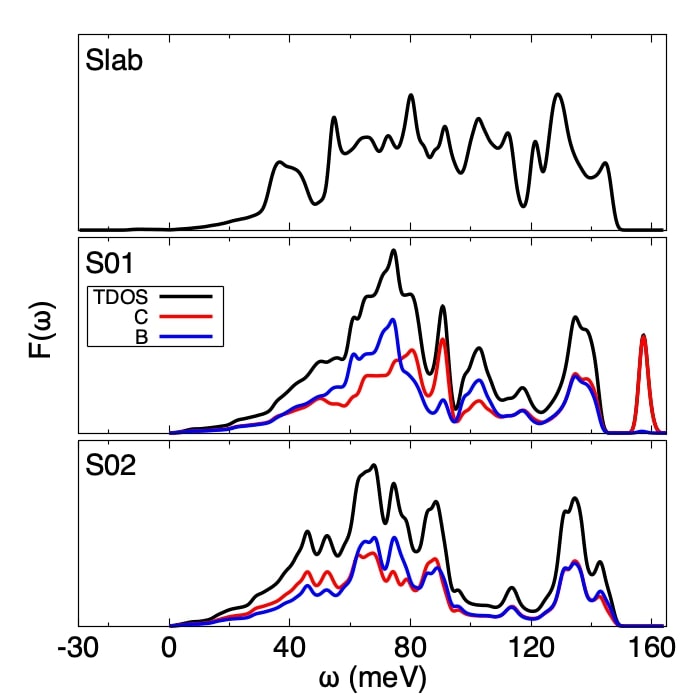}
\caption{The top panel shows the phonon DOS F($\omega$) of the reference slab structure and rest,
the total DOS(black) and partial contribution of C(red) and B(blue) atoms in all the
slab structures.}
\label{fig:SSph}
\end{figure}

\begin{figure}[!htbp]
\includegraphics[width=0.9\columnwidth,angle=0]{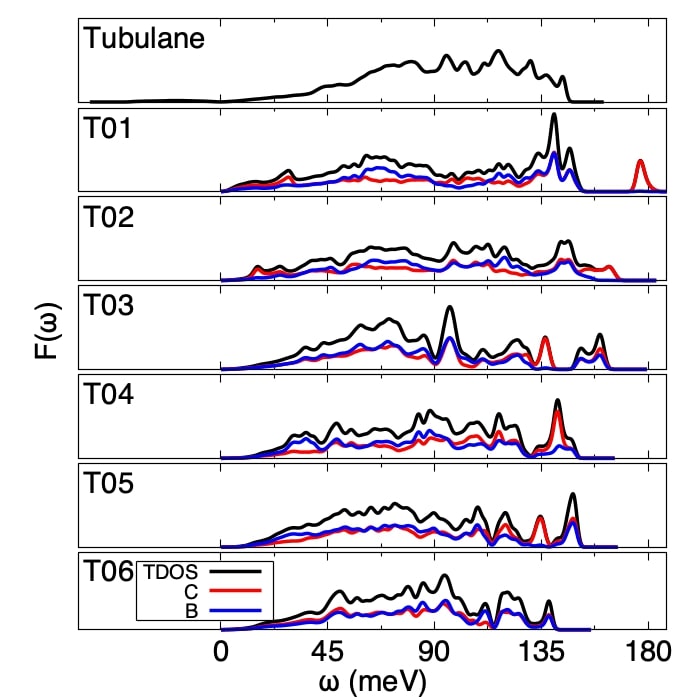}
\caption{The top panel shows the phonon DOS F($\omega$) of the reference tubulane structure and
rest, the total DOS(black) and partial contribution of C(red) and B(blue) atoms in all the
tubulane structures.}
\label{fig:TTph}
\end{figure}

The total(black) and partial phonon DOS(C in red and B in blue) of all the structures arranged 
by family are shown in Fig.\ref{fig:DDph}(D), Fig.\ref{fig:GGph}(G), Fig.\ref{fig:SSph}(S) and
Fig.\ref{fig:TTph}(T).

%\clearpage
%%%%%%%%%%%%%%% SUPERCONDUCTING PROPERTIES
\subsection{Electron-Phonon Spectra}

\begin{figure}[!htbp]
\includegraphics[width=1.00\columnwidth,angle=0]{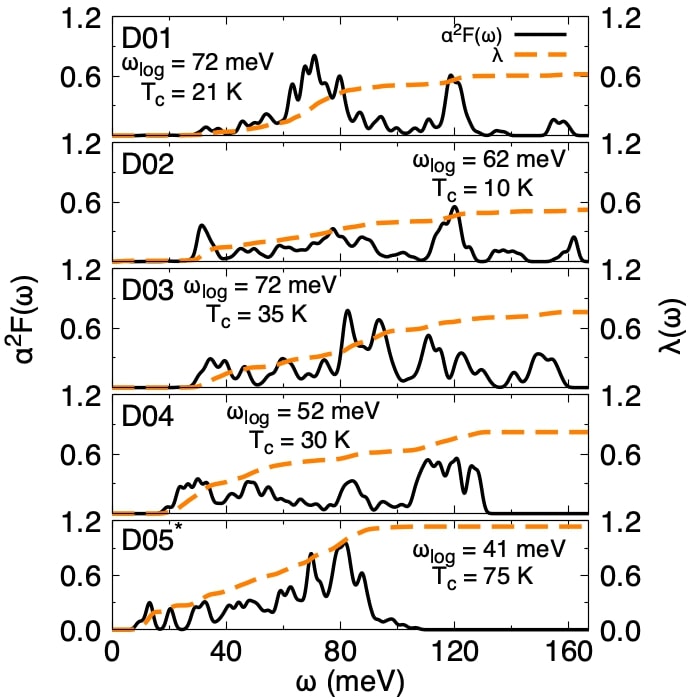}
\caption{The Eliashberg spectral function $\alpha^2$F($\omega$) and \ep\; coupling
constant $\lambda$($\omega$) of the diamond structures. The $\alpha^2$F($\omega$) 
and $\lambda$($\omega$) of D05 is scaled down by 0.5}
\label{fig:DDa2F}
\end{figure}

\begin{figure}[!htb]
\includegraphics[width=1.00\columnwidth,angle=0]{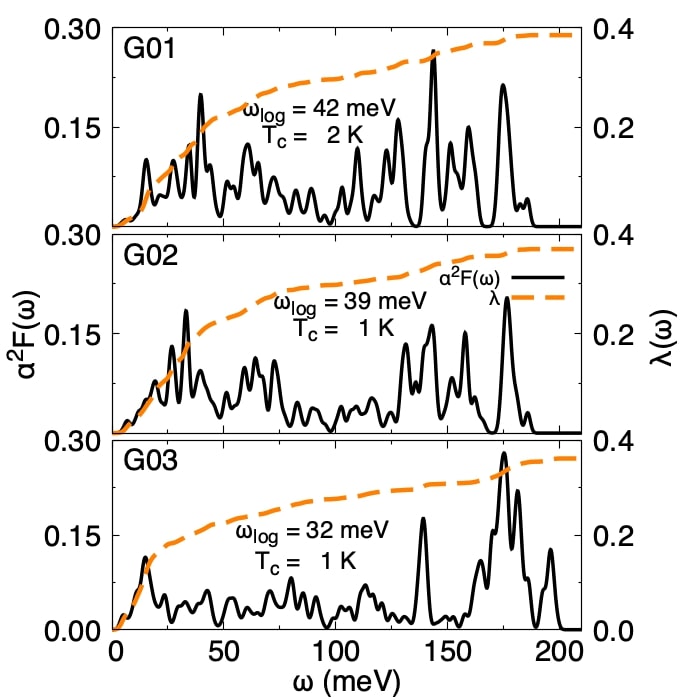}
\caption{The Eliashberg spectral function $\alpha^2$F($\omega$) and \ep\; coupling
constant $\lambda$($\omega$) of the graphite structures.}
\label{fig:GGa2F}
\end{figure}

\begin{figure}[!htb]
\includegraphics[width=1.00\columnwidth,angle=0]{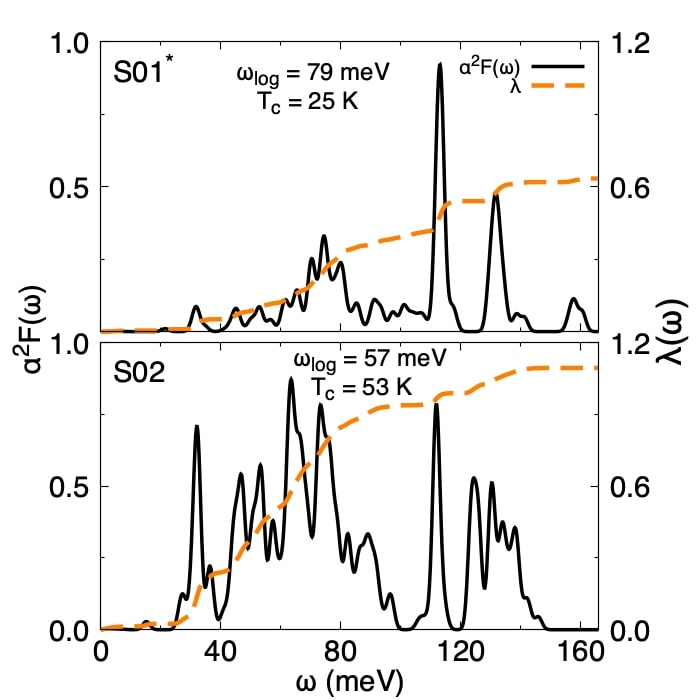}
\caption{The Eliashberg spectral function $\alpha^2$F($\omega$) and \ep\; coupling
constant $\lambda$($\omega$) of the slab structures. The $\alpha^2$F($\omega$) of 
S01 is scaled down by 0.5}
\label{fig:SSa2F}
\end{figure}

\begin{figure}[!htb]
\includegraphics[width=1.00\columnwidth,angle=0]{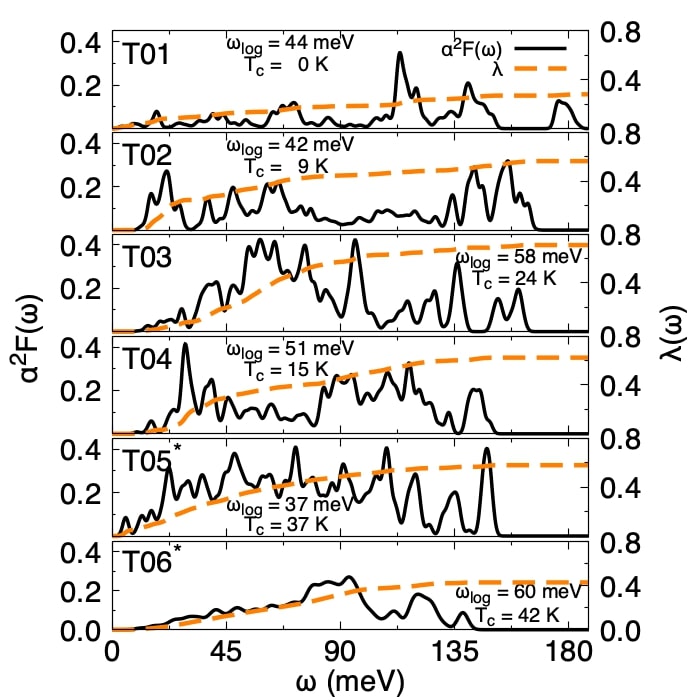}
\caption{The Eliashberg spectral function $\alpha^2$F($\omega$) and \ep\; coupling
constant $\lambda$($\omega$) of the tubulane structures. The $\lambda$($\omega$) of 
T05 is scaled down by 0.5. For T06, both the $\alpha^2$F($\omega$) and
$\lambda$($\omega$) are scaled down by factor of 0.5}
\label{fig:TTa2F}
\end{figure}
%\clearpage

\section{COMPUTATIONAL METHODS}
\subsection{Structure Prediction and DFT Calculations}
%%%MH RUNS AND SETTINGS
The minima hopping (MH) method ~\cite{Goedecker_2004,Goedecker_2005,Amsler_2010,Book:Amsler2018,amsler2019flame} was
used for an efficient scanning of the potential energy surface to find low-energy
structures. 
The  DFT calculations for total-energy and relaxations were carried out using the
Vienna Ab-initio Simulation Package (VASP)  ~\cite{VASP_Kresse,kresse1996efficient}; 
B and C atoms were described by the built-in Projector Augmented Wave (PAW)
potentials\cite{PAW-VASP} with the Perdew-Burke-Ernzerhof (PBE) exchange-correlation
functional ~\cite{GGA-PBE}. The energy cutoff used for the DFT runs was 380 eV, 
which ensures an accuracy of $\sim$ 10 meV/atom.

%%%%RELAXATION RUNS / ENERGY AND ELECTRONIC DOS SETTINGS (PBE-D3)
The post-relaxation, energy evaluation and calculation of the electronic DOS of all
structures were performed using VASP with same set of PAW potentials and PBE 
functional as used in the MH runs, but including van-der-Waals D3 dispersion
corrections with Becke-Jonson damping\cite{D3-BJ}. A threshold of 1 meV/\AA\;
force of each atom and 0.1 KBar on stress was set for the relaxation. The energy 
cutoff used for the post-relaxation calculation was 500 eV. For the relaxation and
energy evaluation, the reciprocal ($\mathbf{k}$) space integration employed
a uniform $\mathbf{k}$-grid with a resolution of 2$\pi$ $\times$ 0.10 \AA$^{-1}$
centered at the $\Gamma$-point a gaussian smearing of width 0.10 eV. For an accurate
evalutation of the electronic DOS we employed the improved Tetrahedron method, as
implemented in VASP.~\cite{DFT:OKA_tetra_PRB1994}

%%%%%%%%%%%PHONON CALCULATION
\subsection{Phonon and Electron-Phonon Coupling Calculations}
The phonon calculations at the $\Gamma$-point on the post-relaxed structures and 
the complete phonon calculations of phonon spectra and \ep\; matrix
elements  were carried out within Density Functional Perturbation Theory(DFPT), as
implemented in the plane-wave pseudopotential code \textit{Quantum Espresso}-6.4.1
\cite{QE-2009,QE-2017}. 
Atoms were described by Optimized Norm-Conserving Vanderbilt  (ONCV) pseudopotentials
\cite{hamann2013optimized}. For the initial phonon calculations at the $\Gamma$-point
on the large database of structures, a PBE functional was used, whereas the remaining
calculations were done with ONCV pseudopotentials\cite{van2018pseudodojo} 
with Perdew-Wang92-LDA functional\cite{LDA-PW}, which ensures more accurate 
relaxations for layered structures.
The \ep\; matrix calculations were carried out on regular $\Gamma$-centered
Monkhorst-Pack(MP)4 $\times$ 4 $\times$ 4 grids for phonons ($\mathbf{q}$) and 8
$\times$ 8 $\times$ 8 grids for electrons ($\mathbf{k}$). The selected
structures were re-relaxed to a threshold force of 0.1 meV/\AA\;  and stress 
of 0.1 KBar prior to phonon and \ep; calculation with the LDA functional.
For these calculations, an energy cutoff of 80 Ry was used with a Gaussian 
smearing of 0.04 Ry for $\mathbf{k}$-space integration.
For all structures, we employed a 8-atoms supercell. In the case of the D05 
structure, where symmetry allowed us to reduce the structure to two atoms/cell, 
a phonon grid of q= 8 $\times$ 8 $\times$ 8 mesh  was used.

\subsection{Screening Protocol and Structure Selection}
In this work, we have developed a three-stage protocol to identify 
superconductivity candidates from an initial {\em pool} of 320 metastable 
MH structures. In the first step, we wanted to prune out structures which had no
potential for superconductivity.
We thus needed to identify structure which should be: (i) plausible to be realized in experimental conditions\cite{aykol2018thermodynamic}, 
(ii) metallic (iii) dynamically stable and
(iv) exhibit stiff directional bonds, which ensure
large phonon frequencies and \ep\; matrix elements.

Each of the these qualitative features can be estimated by the energy of formation
$\Delta$E (i), electronic DOS at the Fermi level N($E_F$) (ii), and the phonon 
spectrum (iii), respectively. The energy of formation and the electronic DOS were
already calculated for all the 320 structures after the post-relaxation step. 

For the dynamical stability, a full calculation of the phonon spectrum is too 
expensive to be feasible, whereas calculating the phonon frequencies $\omega_i$ only
at the $\Gamma$-point is relatively inexpensive. Hence, we calculated the $\omega_i$'s
only at the $\Gamma$-point. This approach is not sufficient to assess the dynamical
stability of the structure but it allowed us to reduce the pool of candidates, 
removing structures with imaginary $\omega_i$'s. 

From the $\omega_i$'s, we further constructed a single-number descriptor
$\omega_{avg}$, which is an average of all optical phonon frequencies at the
$\Gamma$-point.

With these three quantities, i.e. $\Delta$E, N($E_F$) and $\omega_{avg}$, in hand, we then developed and used the following three-steps screening process:
\begin{itemize}
    \item Step 1: Structures with $\Delta E$ $\leqslant$ 0.5 eV/atom w.r.t Graphite and $\alpha$-Rhombohedral-B$_{12}$ are retained.
    \item Step 2: Structures with N($E_F$) $\geqslant$ 0.1 states/eV/atom are retained.
    \item Step 3: Structures with N($E_F$) $\times$ $\omega_{avg}$ $\geqslant$ 0.02 states/atom are retained.
\end{itemize}
These requirements are still broad enough that the initial subset was reduced to 116
candidates, which still constitute a too large pool for a complete calculation of
superconducting properties. To reduce the total pool to a manageable number, we 
hand-picked a few structures (5-6) for each family (D, G, S and T) from the final set 
of screened candidates. While selecting the structures, care was taken that they
represent different values of N($E_F$) and $\omega_{avg}$ values. For these 
structures, we performed full calculations of the phonon  and \ep\; coupling spectra
over the full BZ. If any of the initally-selected structures was thus found to be
dynamically unstable, it was replaced with another candidate with similar
values of N($E_F$) and $\omega_{avg}$. The positions of the selected structures in
the energy vs volume plot in Fig.\ref{fig:volene} are shown by the dark colored
symbols, whereas the rest of the structures are indicated by the light colored
symbols.

\subsection{Convex Hull}
\begin{figure}[!htbp]
\includegraphics[width=1.0\columnwidth,angle=0]{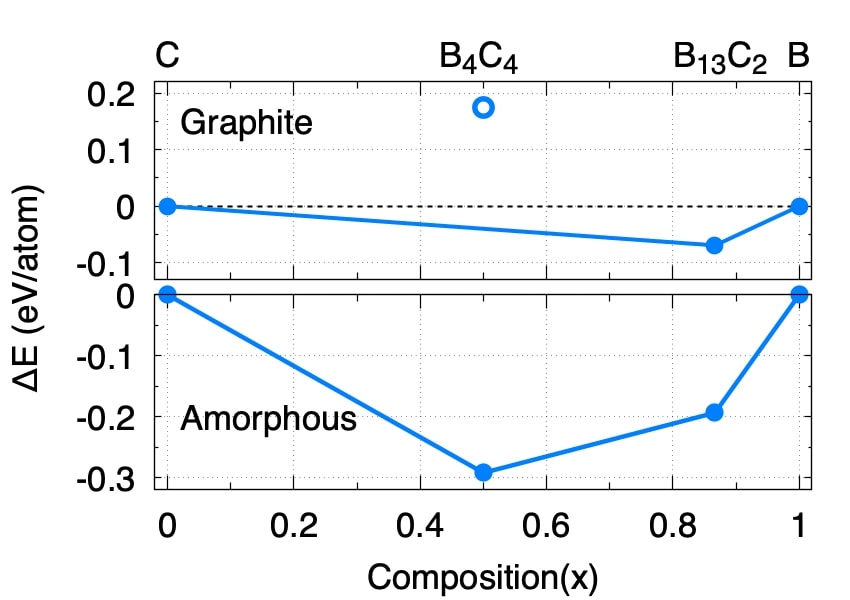}
\caption{Convex hull of the Boron-Carbon system w.r.t. graphite C and 
amorphous C in top and bottom panel respectively. The reference for B
is $\alpha$-R-B$_{12}$. The reference energy for the amorphous
C is obtained from Ref.~\onlinecite{aykol2018thermodynamic}.}
\label{fig:Convexhull}

\end{figure}
\bibliographystyle{apsrev4-1}
\bibliography{references}

%merlin.mbs apsrev4-1.bst 2010-07-25 4.21a (PWD, AO, DPC) hacked
%Control: key (0)
%Control: author (72) initials jnrlst
%Control: editor formatted (1) identically to author
%Control: production of article title (-1) disabled
%Control: page (0) single
%Control: year (1) truncated
%Control: production of eprint (0) enabled
\begin{thebibliography}{66}%
\makeatletter
\providecommand \@ifxundefined [1]{%
 \@ifx{#1\undefined}
}%
\providecommand \@ifnum [1]{%
 \ifnum #1\expandafter \@firstoftwo
 \else \expandafter \@secondoftwo
 \fi
}%
\providecommand \@ifx [1]{%
 \ifx #1\expandafter \@firstoftwo
 \else \expandafter \@secondoftwo
 \fi
}%
\providecommand \natexlab [1]{#1}%
\providecommand \enquote  [1]{``#1''}%
\providecommand \bibnamefont  [1]{#1}%
\providecommand \bibfnamefont [1]{#1}%
\providecommand \citenamefont [1]{#1}%
\providecommand \href@noop [0]{\@secondoftwo}%
\providecommand \href [0]{\begingroup \@sanitize@url \@href}%
\providecommand \@href[1]{\@@startlink{#1}\@@href}%
\providecommand \@@href[1]{\endgroup#1\@@endlink}%
\providecommand \@sanitize@url [0]{\catcode `\\12\catcode `\$12\catcode
  `\&12\catcode `\#12\catcode `\^12\catcode `\_12\catcode `\%12\relax}%
\providecommand \@@startlink[1]{}%
\providecommand \@@endlink[0]{}%
\providecommand \url  [0]{\begingroup\@sanitize@url \@url }%
\providecommand \@url [1]{\endgroup\@href {#1}{\urlprefix }}%
\providecommand \urlprefix  [0]{URL }%
\providecommand \Eprint [0]{\href }%
\providecommand \doibase [0]{http://dx.doi.org/}%
\providecommand \selectlanguage [0]{\@gobble}%
\providecommand \bibinfo  [0]{\@secondoftwo}%
\providecommand \bibfield  [0]{\@secondoftwo}%
\providecommand \translation [1]{[#1]}%
\providecommand \BibitemOpen [0]{}%
\providecommand \bibitemStop [0]{}%
\providecommand \bibitemNoStop [0]{.\EOS\space}%
\providecommand \EOS [0]{\spacefactor3000\relax}%
\providecommand \BibitemShut  [1]{\csname bibitem#1\endcsname}%
\let\auto@bib@innerbib\@empty
%</preamble>
\bibitem [{\citenamefont {Ashcroft}(1968)}]{ashcroft1968metallic}%
  \BibitemOpen
  \bibfield  {author} {\bibinfo {author} {\bibfnamefont {N.~W.}\ \bibnamefont
  {Ashcroft}},\ }\href@noop {} {\bibfield  {journal} {\bibinfo  {journal}
  {Physical Review Letters}\ }\textbf {\bibinfo {volume} {21}},\ \bibinfo
  {pages} {1748} (\bibinfo {year} {1968})}\BibitemShut {NoStop}%
\bibitem [{\citenamefont {Wigner}\ and\ \citenamefont
  {Huntington}(1935)}]{Huntington1935}%
  \BibitemOpen
  \bibfield  {author} {\bibinfo {author} {\bibfnamefont {E.}~\bibnamefont
  {Wigner}}\ and\ \bibinfo {author} {\bibfnamefont {H.~B.}\ \bibnamefont
  {Huntington}},\ }\href {\doibase 10.1063/1.1749590} {\bibfield  {journal}
  {\bibinfo  {journal} {The Journal of Chemical Physics}\ }\textbf {\bibinfo
  {volume} {3}},\ \bibinfo {pages} {764} (\bibinfo {year} {1935})},\ \Eprint
  {http://arxiv.org/abs/https://doi.org/10.1063/1.1749590}
  {https://doi.org/10.1063/1.1749590} \BibitemShut {NoStop}%
\bibitem [{\citenamefont {Flores-Livas}\ \emph {et~al.}(2020)\citenamefont
  {Flores-Livas}, \citenamefont {Boeri}, \citenamefont {Sanna}, \citenamefont
  {Profeta}, \citenamefont {Arita},\ and\ \citenamefont
  {Eremets}}]{Hydrides:our_review}%
  \BibitemOpen
  \bibfield  {author} {\bibinfo {author} {\bibfnamefont {J.~A.}\ \bibnamefont
  {Flores-Livas}}, \bibinfo {author} {\bibfnamefont {L.}~\bibnamefont {Boeri}},
  \bibinfo {author} {\bibfnamefont {A.}~\bibnamefont {Sanna}}, \bibinfo
  {author} {\bibfnamefont {G.}~\bibnamefont {Profeta}}, \bibinfo {author}
  {\bibfnamefont {R.}~\bibnamefont {Arita}}, \ and\ \bibinfo {author}
  {\bibfnamefont {M.}~\bibnamefont {Eremets}},\ }\href {\doibase
  https://doi.org/10.1016/j.physrep.2020.02.003} {\bibfield  {journal}
  {\bibinfo  {journal} {Physics Reports}\ } (\bibinfo {year} {2020}),\
  https://doi.org/10.1016/j.physrep.2020.02.003}\BibitemShut {NoStop}%
\bibitem [{\citenamefont {Eremets}\ \emph {et~al.}(2019)\citenamefont
  {Eremets}, \citenamefont {Drozdov}, \citenamefont {Kong},\ and\ \citenamefont
  {Wang}}]{H:Eremets_2019}%
  \BibitemOpen
  \bibfield  {author} {\bibinfo {author} {\bibfnamefont {M.~I.}\ \bibnamefont
  {Eremets}}, \bibinfo {author} {\bibfnamefont {A.~P.}\ \bibnamefont
  {Drozdov}}, \bibinfo {author} {\bibfnamefont {P.~P.}\ \bibnamefont {Kong}}, \
  and\ \bibinfo {author} {\bibfnamefont {H.}~\bibnamefont {Wang}},\ }\href
  {\doibase 10.1038/s41567-019-0646-x} {\bibfield  {journal} {\bibinfo
  {journal} {Nature Physics}\ }\textbf {\bibinfo {volume} {15}},\ \bibinfo
  {pages} {1246} (\bibinfo {year} {2019})}\BibitemShut {NoStop}%
\bibitem [{\citenamefont {Loubeyre}\ \emph {et~al.}(2020)\citenamefont
  {Loubeyre}, \citenamefont {Occelli},\ and\ \citenamefont
  {Dumas}}]{H:Loubeyre_2020}%
  \BibitemOpen
  \bibfield  {author} {\bibinfo {author} {\bibfnamefont {P.}~\bibnamefont
  {Loubeyre}}, \bibinfo {author} {\bibfnamefont {F.}~\bibnamefont {Occelli}}, \
  and\ \bibinfo {author} {\bibfnamefont {P.}~\bibnamefont {Dumas}},\ }\href
  {\doibase 10.1038/s41586-019-1927-3} {\bibfield  {journal} {\bibinfo
  {journal} {Nature}\ }\textbf {\bibinfo {volume} {577}},\ \bibinfo {pages}
  {631} (\bibinfo {year} {2020})}\BibitemShut {NoStop}%
\bibitem [{\citenamefont {Ashcroft}(2004)}]{PhysRevLett.92.187002}%
  \BibitemOpen
  \bibfield  {author} {\bibinfo {author} {\bibfnamefont {N.~W.}\ \bibnamefont
  {Ashcroft}},\ }\href {\doibase 10.1103/PhysRevLett.92.187002} {\bibfield
  {journal} {\bibinfo  {journal} {Phys. Rev. Lett.}\ }\textbf {\bibinfo
  {volume} {92}},\ \bibinfo {pages} {187002} (\bibinfo {year}
  {2004})}\BibitemShut {NoStop}%
\bibitem [{\citenamefont {Duan}\ \emph {et~al.}(2014)\citenamefont {Duan},
  \citenamefont {Liu}, \citenamefont {Tian}, \citenamefont {Li}, \citenamefont
  {Huang}, \citenamefont {Zhao}, \citenamefont {Yu}, \citenamefont {Liu},
  \citenamefont {Tian},\ and\ \citenamefont {Cui}}]{duan2014pressure}%
  \BibitemOpen
  \bibfield  {author} {\bibinfo {author} {\bibfnamefont {D.}~\bibnamefont
  {Duan}}, \bibinfo {author} {\bibfnamefont {Y.}~\bibnamefont {Liu}}, \bibinfo
  {author} {\bibfnamefont {F.}~\bibnamefont {Tian}}, \bibinfo {author}
  {\bibfnamefont {D.}~\bibnamefont {Li}}, \bibinfo {author} {\bibfnamefont
  {X.}~\bibnamefont {Huang}}, \bibinfo {author} {\bibfnamefont
  {Z.}~\bibnamefont {Zhao}}, \bibinfo {author} {\bibfnamefont {H.}~\bibnamefont
  {Yu}}, \bibinfo {author} {\bibfnamefont {B.}~\bibnamefont {Liu}}, \bibinfo
  {author} {\bibfnamefont {W.}~\bibnamefont {Tian}}, \ and\ \bibinfo {author}
  {\bibfnamefont {T.}~\bibnamefont {Cui}},\ }\href@noop {} {\bibfield
  {journal} {\bibinfo  {journal} {Scientific reports}\ }\textbf {\bibinfo
  {volume} {4}},\ \bibinfo {pages} {6968} (\bibinfo {year} {2014})}\BibitemShut
  {NoStop}%
\bibitem [{\citenamefont {Drozdov}\ \emph {et~al.}(2015)\citenamefont
  {Drozdov}, \citenamefont {Eremets},\ and\ \citenamefont
  {Troyan}}]{drozdov2015superconductivity}%
  \BibitemOpen
  \bibfield  {author} {\bibinfo {author} {\bibfnamefont {A.}~\bibnamefont
  {Drozdov}}, \bibinfo {author} {\bibfnamefont {M.}~\bibnamefont {Eremets}}, \
  and\ \bibinfo {author} {\bibfnamefont {I.}~\bibnamefont {Troyan}},\
  }\href@noop {} {\bibfield  {journal} {\bibinfo  {journal} {arXiv preprint
  arXiv:1508.06224}\ } (\bibinfo {year} {2015})}\BibitemShut {NoStop}%
\bibitem [{\citenamefont {Liu}\ \emph {et~al.}(2017)\citenamefont {Liu},
  \citenamefont {Naumov}, \citenamefont {Hoffmann}, \citenamefont {Ashcroft},\
  and\ \citenamefont {Hemley}}]{liu2017potential}%
  \BibitemOpen
  \bibfield  {author} {\bibinfo {author} {\bibfnamefont {H.}~\bibnamefont
  {Liu}}, \bibinfo {author} {\bibfnamefont {I.~I.}\ \bibnamefont {Naumov}},
  \bibinfo {author} {\bibfnamefont {R.}~\bibnamefont {Hoffmann}}, \bibinfo
  {author} {\bibfnamefont {N.}~\bibnamefont {Ashcroft}}, \ and\ \bibinfo
  {author} {\bibfnamefont {R.~J.}\ \bibnamefont {Hemley}},\ }\href@noop {}
  {\bibfield  {journal} {\bibinfo  {journal} {Proceedings of the National
  Academy of Sciences}\ }\textbf {\bibinfo {volume} {114}},\ \bibinfo {pages}
  {6990} (\bibinfo {year} {2017})}\BibitemShut {NoStop}%
\bibitem [{\citenamefont {Somayazulu}\ \emph {et~al.}(2019)\citenamefont
  {Somayazulu}, \citenamefont {Ahart}, \citenamefont {Mishra}, \citenamefont
  {Geballe}, \citenamefont {Baldini}, \citenamefont {Meng}, \citenamefont
  {Struzhkin},\ and\ \citenamefont {Hemley}}]{LaH10:hemley_PRL2019}%
  \BibitemOpen
  \bibfield  {author} {\bibinfo {author} {\bibfnamefont {M.}~\bibnamefont
  {Somayazulu}}, \bibinfo {author} {\bibfnamefont {M.}~\bibnamefont {Ahart}},
  \bibinfo {author} {\bibfnamefont {A.~K.}\ \bibnamefont {Mishra}}, \bibinfo
  {author} {\bibfnamefont {Z.~M.}\ \bibnamefont {Geballe}}, \bibinfo {author}
  {\bibfnamefont {M.}~\bibnamefont {Baldini}}, \bibinfo {author} {\bibfnamefont
  {Y.}~\bibnamefont {Meng}}, \bibinfo {author} {\bibfnamefont {V.~V.}\
  \bibnamefont {Struzhkin}}, \ and\ \bibinfo {author} {\bibfnamefont {R.~J.}\
  \bibnamefont {Hemley}},\ }\href {\doibase 10.1103/PhysRevLett.122.027001}
  {\bibfield  {journal} {\bibinfo  {journal} {Phys. Rev. Lett.}\ }\textbf
  {\bibinfo {volume} {122}},\ \bibinfo {pages} {027001} (\bibinfo {year}
  {2019})}\BibitemShut {NoStop}%
\bibitem [{\citenamefont {Drozdov}\ \emph {et~al.}(2019)\citenamefont
  {Drozdov}, \citenamefont {Kong}, \citenamefont {Minkov}, \citenamefont
  {Besedin}, \citenamefont {Kuzovnikov}, \citenamefont {Mozaffari},
  \citenamefont {Balicas}, \citenamefont {Balakirev}, \citenamefont {Graf},
  \citenamefont {Prakapenka} \emph {et~al.}}]{drozdov2019superconductivity}%
  \BibitemOpen
  \bibfield  {author} {\bibinfo {author} {\bibfnamefont {A.}~\bibnamefont
  {Drozdov}}, \bibinfo {author} {\bibfnamefont {P.}~\bibnamefont {Kong}},
  \bibinfo {author} {\bibfnamefont {V.}~\bibnamefont {Minkov}}, \bibinfo
  {author} {\bibfnamefont {S.}~\bibnamefont {Besedin}}, \bibinfo {author}
  {\bibfnamefont {M.}~\bibnamefont {Kuzovnikov}}, \bibinfo {author}
  {\bibfnamefont {S.}~\bibnamefont {Mozaffari}}, \bibinfo {author}
  {\bibfnamefont {L.}~\bibnamefont {Balicas}}, \bibinfo {author} {\bibfnamefont
  {F.}~\bibnamefont {Balakirev}}, \bibinfo {author} {\bibfnamefont
  {D.}~\bibnamefont {Graf}}, \bibinfo {author} {\bibfnamefont {V.}~\bibnamefont
  {Prakapenka}},  \emph {et~al.},\ }\href@noop {} {\bibfield  {journal}
  {\bibinfo  {journal} {Nature}\ }\textbf {\bibinfo {volume} {569}},\ \bibinfo
  {pages} {528} (\bibinfo {year} {2019})}\BibitemShut {NoStop}%
\bibitem [{\citenamefont {Boeri}\ and\ \citenamefont
  {Bachelet}(2019)}]{SC:Boeri_Bachelet_JPCM2019}%
  \BibitemOpen
  \bibfield  {author} {\bibinfo {author} {\bibfnamefont {L.}~\bibnamefont
  {Boeri}}\ and\ \bibinfo {author} {\bibfnamefont {G.~B.}\ \bibnamefont
  {Bachelet}},\ }\href {\doibase 10.1088/1361-648x/ab0db2} {\bibfield
  {journal} {\bibinfo  {journal} {Journal of Physics: Condensed Matter}\
  }\textbf {\bibinfo {volume} {31}},\ \bibinfo {pages} {234002} (\bibinfo
  {year} {2019})}\BibitemShut {NoStop}%
\bibitem [{\citenamefont {Nagamatsu}\ \emph {et~al.}(2001)\citenamefont
  {Nagamatsu}, \citenamefont {Nakagawa}, \citenamefont {Muranaka},
  \citenamefont {Zenitani},\ and\ \citenamefont
  {Akimitsu}}]{nagamatsu2001superconductivity}%
  \BibitemOpen
  \bibfield  {author} {\bibinfo {author} {\bibfnamefont {J.}~\bibnamefont
  {Nagamatsu}}, \bibinfo {author} {\bibfnamefont {N.}~\bibnamefont {Nakagawa}},
  \bibinfo {author} {\bibfnamefont {T.}~\bibnamefont {Muranaka}}, \bibinfo
  {author} {\bibfnamefont {Y.}~\bibnamefont {Zenitani}}, \ and\ \bibinfo
  {author} {\bibfnamefont {J.}~\bibnamefont {Akimitsu}},\ }\href@noop {}
  {\bibfield  {journal} {\bibinfo  {journal} {nature}\ }\textbf {\bibinfo
  {volume} {410}},\ \bibinfo {pages} {63} (\bibinfo {year} {2001})}\BibitemShut
  {NoStop}%
\bibitem [{\citenamefont {An}\ and\ \citenamefont
  {Pickett}(2001)}]{SC:mgb2_Pickett_PRL2001}%
  \BibitemOpen
  \bibfield  {author} {\bibinfo {author} {\bibfnamefont {J.~M.}\ \bibnamefont
  {An}}\ and\ \bibinfo {author} {\bibfnamefont {W.~E.}\ \bibnamefont
  {Pickett}},\ }\href {\doibase 10.1103/PhysRevLett.86.4366} {\bibfield
  {journal} {\bibinfo  {journal} {Phys. Rev. Lett.}\ }\textbf {\bibinfo
  {volume} {86}},\ \bibinfo {pages} {4366} (\bibinfo {year}
  {2001})}\BibitemShut {NoStop}%
\bibitem [{\citenamefont {Ekimov}\ \emph {et~al.}(2004)\citenamefont {Ekimov},
  \citenamefont {Sidorov}, \citenamefont {Bauer}, \citenamefont {Mel'Nik},
  \citenamefont {Curro}, \citenamefont {Thompson},\ and\ \citenamefont
  {Stishov}}]{ekimov2004superconductivity}%
  \BibitemOpen
  \bibfield  {author} {\bibinfo {author} {\bibfnamefont {E.}~\bibnamefont
  {Ekimov}}, \bibinfo {author} {\bibfnamefont {V.}~\bibnamefont {Sidorov}},
  \bibinfo {author} {\bibfnamefont {E.}~\bibnamefont {Bauer}}, \bibinfo
  {author} {\bibfnamefont {N.}~\bibnamefont {Mel'Nik}}, \bibinfo {author}
  {\bibfnamefont {N.}~\bibnamefont {Curro}}, \bibinfo {author} {\bibfnamefont
  {J.}~\bibnamefont {Thompson}}, \ and\ \bibinfo {author} {\bibfnamefont
  {S.}~\bibnamefont {Stishov}},\ }\href@noop {} {\bibfield  {journal} {\bibinfo
   {journal} {nature}\ }\textbf {\bibinfo {volume} {428}},\ \bibinfo {pages}
  {542} (\bibinfo {year} {2004})}\BibitemShut {NoStop}%
\bibitem [{\citenamefont {Boeri}\ \emph {et~al.}(2004)\citenamefont {Boeri},
  \citenamefont {Kortus},\ and\ \citenamefont
  {Andersen}}]{SC:diamond_Boeri_PRL2004}%
  \BibitemOpen
  \bibfield  {author} {\bibinfo {author} {\bibfnamefont {L.}~\bibnamefont
  {Boeri}}, \bibinfo {author} {\bibfnamefont {J.}~\bibnamefont {Kortus}}, \
  and\ \bibinfo {author} {\bibfnamefont {O.~K.}\ \bibnamefont {Andersen}},\
  }\href {\doibase 10.1103/PhysRevLett.93.237002} {\bibfield  {journal}
  {\bibinfo  {journal} {Phys. Rev. Lett.}\ }\textbf {\bibinfo {volume} {93}},\
  \bibinfo {pages} {237002} (\bibinfo {year} {2004})}\BibitemShut {NoStop}%
\bibitem [{\citenamefont {Lee}\ and\ \citenamefont
  {Pickett}(2004)}]{SC:diamond_Pickett_PRL2004}%
  \BibitemOpen
  \bibfield  {author} {\bibinfo {author} {\bibfnamefont {K.-W.}\ \bibnamefont
  {Lee}}\ and\ \bibinfo {author} {\bibfnamefont {W.~E.}\ \bibnamefont
  {Pickett}},\ }\href@noop {} {\bibfield  {journal} {\bibinfo  {journal}
  {Physical review letters}\ }\textbf {\bibinfo {volume} {93}},\ \bibinfo
  {pages} {237003} (\bibinfo {year} {2004})}\BibitemShut {NoStop}%
\bibitem [{\citenamefont {Blase}\ \emph {et~al.}(2004)\citenamefont {Blase},
  \citenamefont {Adessi},\ and\ \citenamefont
  {Conn\'etable}}]{SC:Blase_PRL_2004}%
  \BibitemOpen
  \bibfield  {author} {\bibinfo {author} {\bibfnamefont {X.}~\bibnamefont
  {Blase}}, \bibinfo {author} {\bibfnamefont {C.}~\bibnamefont {Adessi}}, \
  and\ \bibinfo {author} {\bibfnamefont {D.}~\bibnamefont {Conn\'etable}},\
  }\href {\doibase 10.1103/PhysRevLett.93.237004} {\bibfield  {journal}
  {\bibinfo  {journal} {Phys. Rev. Lett.}\ }\textbf {\bibinfo {volume} {93}},\
  \bibinfo {pages} {237004} (\bibinfo {year} {2004})}\BibitemShut {NoStop}%
\bibitem [{\citenamefont {Giustino}\ \emph {et~al.}(2007)\citenamefont
  {Giustino}, \citenamefont {Yates}, \citenamefont {Souza}, \citenamefont
  {Cohen},\ and\ \citenamefont {Louie}}]{SC:diamond_giustino_PRL_2007}%
  \BibitemOpen
  \bibfield  {author} {\bibinfo {author} {\bibfnamefont {F.}~\bibnamefont
  {Giustino}}, \bibinfo {author} {\bibfnamefont {J.~R.}\ \bibnamefont {Yates}},
  \bibinfo {author} {\bibfnamefont {I.}~\bibnamefont {Souza}}, \bibinfo
  {author} {\bibfnamefont {M.~L.}\ \bibnamefont {Cohen}}, \ and\ \bibinfo
  {author} {\bibfnamefont {S.~G.}\ \bibnamefont {Louie}},\ }\href {\doibase
  10.1103/PhysRevLett.98.047005} {\bibfield  {journal} {\bibinfo  {journal}
  {Phys. Rev. Lett.}\ }\textbf {\bibinfo {volume} {98}},\ \bibinfo {pages}
  {047005} (\bibinfo {year} {2007})}\BibitemShut {NoStop}%
\bibitem [{\citenamefont {Rosner}\ \emph {et~al.}(2002)\citenamefont {Rosner},
  \citenamefont {Kitaigorodsky},\ and\ \citenamefont
  {Pickett}}]{SC:Rosner_LiBC_PRL2002}%
  \BibitemOpen
  \bibfield  {author} {\bibinfo {author} {\bibfnamefont {H.}~\bibnamefont
  {Rosner}}, \bibinfo {author} {\bibfnamefont {A.}~\bibnamefont
  {Kitaigorodsky}}, \ and\ \bibinfo {author} {\bibfnamefont {W.~E.}\
  \bibnamefont {Pickett}},\ }\href {\doibase 10.1103/PhysRevLett.88.127001}
  {\bibfield  {journal} {\bibinfo  {journal} {Phys. Rev. Lett.}\ }\textbf
  {\bibinfo {volume} {88}},\ \bibinfo {pages} {127001} (\bibinfo {year}
  {2002})}\BibitemShut {NoStop}%
\bibitem [{\citenamefont {Kolmogorov}\ and\ \citenamefont
  {Curtarolo}(2006)}]{SC:LIB_Kolmogorov_PRB2006}%
  \BibitemOpen
  \bibfield  {author} {\bibinfo {author} {\bibfnamefont {A.~N.}\ \bibnamefont
  {Kolmogorov}}\ and\ \bibinfo {author} {\bibfnamefont {S.}~\bibnamefont
  {Curtarolo}},\ }\href {\doibase 10.1103/PhysRevB.73.180501} {\bibfield
  {journal} {\bibinfo  {journal} {Phys. Rev. B}\ }\textbf {\bibinfo {volume}
  {73}},\ \bibinfo {pages} {180501} (\bibinfo {year} {2006})}\BibitemShut
  {NoStop}%
\bibitem [{\citenamefont {Savini}\ \emph {et~al.}(2010)\citenamefont {Savini},
  \citenamefont {Ferrari},\ and\ \citenamefont
  {Giustino}}]{SC:graphane_savini_PRL2010}%
  \BibitemOpen
  \bibfield  {author} {\bibinfo {author} {\bibfnamefont {G.}~\bibnamefont
  {Savini}}, \bibinfo {author} {\bibfnamefont {A.~C.}\ \bibnamefont {Ferrari}},
  \ and\ \bibinfo {author} {\bibfnamefont {F.}~\bibnamefont {Giustino}},\
  }\href {\doibase 10.1103/PhysRevLett.105.037002} {\bibfield  {journal}
  {\bibinfo  {journal} {Phys. Rev. Lett.}\ }\textbf {\bibinfo {volume} {105}},\
  \bibinfo {pages} {037002} (\bibinfo {year} {2010})}\BibitemShut {NoStop}%
\bibitem [{\citenamefont {Alberi}\ \emph {et~al.}(2018)\citenamefont {Alberi},
  \citenamefont {Nardelli}, \citenamefont {Zakutayev}, \citenamefont {Mitas},
  \citenamefont {Curtarolo}, \citenamefont {Jain}, \citenamefont {Fornari},
  \citenamefont {Marzari}, \citenamefont {Takeuchi}, \citenamefont {Green},
  \citenamefont {Kanatzidis}, \citenamefont {Toney}, \citenamefont {Butenko},
  \citenamefont {Meredig}, \citenamefont {Lany}, \citenamefont {Kattner},
  \citenamefont {Davydov}, \citenamefont {Toberer}, \citenamefont {Stevanovic},
  \citenamefont {Walsh}, \citenamefont {Park}, \citenamefont {Aspuru-Guzik},
  \citenamefont {Tabor}, \citenamefont {Nelson}, \citenamefont {Murphy},
  \citenamefont {Setlur}, \citenamefont {Gregoire}, \citenamefont {Li},
  \citenamefont {Xiao}, \citenamefont {Ludwig}, \citenamefont {Martin},
  \citenamefont {Rappe}, \citenamefont {Wei},\ and\ \citenamefont
  {Perkins}}]{ML:MLroadmap_Alberi_2018}%
  \BibitemOpen
  \bibfield  {author} {\bibinfo {author} {\bibfnamefont {K.}~\bibnamefont
  {Alberi}}, \bibinfo {author} {\bibfnamefont {M.~B.}\ \bibnamefont
  {Nardelli}}, \bibinfo {author} {\bibfnamefont {A.}~\bibnamefont {Zakutayev}},
  \bibinfo {author} {\bibfnamefont {L.}~\bibnamefont {Mitas}}, \bibinfo
  {author} {\bibfnamefont {S.}~\bibnamefont {Curtarolo}}, \bibinfo {author}
  {\bibfnamefont {A.}~\bibnamefont {Jain}}, \bibinfo {author} {\bibfnamefont
  {M.}~\bibnamefont {Fornari}}, \bibinfo {author} {\bibfnamefont
  {N.}~\bibnamefont {Marzari}}, \bibinfo {author} {\bibfnamefont
  {I.}~\bibnamefont {Takeuchi}}, \bibinfo {author} {\bibfnamefont {M.~L.}\
  \bibnamefont {Green}}, \bibinfo {author} {\bibfnamefont {M.}~\bibnamefont
  {Kanatzidis}}, \bibinfo {author} {\bibfnamefont {M.~F.}\ \bibnamefont
  {Toney}}, \bibinfo {author} {\bibfnamefont {S.}~\bibnamefont {Butenko}},
  \bibinfo {author} {\bibfnamefont {B.}~\bibnamefont {Meredig}}, \bibinfo
  {author} {\bibfnamefont {S.}~\bibnamefont {Lany}}, \bibinfo {author}
  {\bibfnamefont {U.}~\bibnamefont {Kattner}}, \bibinfo {author} {\bibfnamefont
  {A.}~\bibnamefont {Davydov}}, \bibinfo {author} {\bibfnamefont {E.~S.}\
  \bibnamefont {Toberer}}, \bibinfo {author} {\bibfnamefont {V.}~\bibnamefont
  {Stevanovic}}, \bibinfo {author} {\bibfnamefont {A.}~\bibnamefont {Walsh}},
  \bibinfo {author} {\bibfnamefont {N.-G.}\ \bibnamefont {Park}}, \bibinfo
  {author} {\bibfnamefont {A.}~\bibnamefont {Aspuru-Guzik}}, \bibinfo {author}
  {\bibfnamefont {D.~P.}\ \bibnamefont {Tabor}}, \bibinfo {author}
  {\bibfnamefont {J.}~\bibnamefont {Nelson}}, \bibinfo {author} {\bibfnamefont
  {J.}~\bibnamefont {Murphy}}, \bibinfo {author} {\bibfnamefont
  {A.}~\bibnamefont {Setlur}}, \bibinfo {author} {\bibfnamefont
  {J.}~\bibnamefont {Gregoire}}, \bibinfo {author} {\bibfnamefont
  {H.}~\bibnamefont {Li}}, \bibinfo {author} {\bibfnamefont {R.}~\bibnamefont
  {Xiao}}, \bibinfo {author} {\bibfnamefont {A.}~\bibnamefont {Ludwig}},
  \bibinfo {author} {\bibfnamefont {L.~W.}\ \bibnamefont {Martin}}, \bibinfo
  {author} {\bibfnamefont {A.~M.}\ \bibnamefont {Rappe}}, \bibinfo {author}
  {\bibfnamefont {S.-H.}\ \bibnamefont {Wei}}, \ and\ \bibinfo {author}
  {\bibfnamefont {J.}~\bibnamefont {Perkins}},\ }\href {\doibase
  10.1088/1361-6463/aad926} {\bibfield  {journal} {\bibinfo  {journal} {Journal
  of Physics D: Applied Physics}\ }\textbf {\bibinfo {volume} {52}},\ \bibinfo
  {pages} {013001} (\bibinfo {year} {2018})}\BibitemShut {NoStop}%
\bibitem [{\citenamefont {Schleder}\ \emph {et~al.}(2019)\citenamefont
  {Schleder}, \citenamefont {Padilha}, \citenamefont {Acosta}, \citenamefont
  {Costa},\ and\ \citenamefont {Fazzio}}]{Schleder_MLreview_2019}%
  \BibitemOpen
  \bibfield  {author} {\bibinfo {author} {\bibfnamefont {G.~R.}\ \bibnamefont
  {Schleder}}, \bibinfo {author} {\bibfnamefont {A.~C.~M.}\ \bibnamefont
  {Padilha}}, \bibinfo {author} {\bibfnamefont {C.~M.}\ \bibnamefont {Acosta}},
  \bibinfo {author} {\bibfnamefont {M.}~\bibnamefont {Costa}}, \ and\ \bibinfo
  {author} {\bibfnamefont {A.}~\bibnamefont {Fazzio}},\ }\href {\doibase
  10.1088/2515-7639/ab084b} {\bibfield  {journal} {\bibinfo  {journal} {Journal
  of Physics: Materials}\ }\textbf {\bibinfo {volume} {2}},\ \bibinfo {pages}
  {032001} (\bibinfo {year} {2019})}\BibitemShut {NoStop}%
\bibitem [{\citenamefont {Stanev}\ \emph {et~al.}(2018)\citenamefont {Stanev},
  \citenamefont {Oses}, \citenamefont {Kusne}, \citenamefont {Rodriguez},
  \citenamefont {Paglione}, \citenamefont {Curtarolo},\ and\ \citenamefont
  {Takeuchi}}]{ML:Stanev_TCML_2018}%
  \BibitemOpen
  \bibfield  {author} {\bibinfo {author} {\bibfnamefont {V.}~\bibnamefont
  {Stanev}}, \bibinfo {author} {\bibfnamefont {C.}~\bibnamefont {Oses}},
  \bibinfo {author} {\bibfnamefont {A.~G.}\ \bibnamefont {Kusne}}, \bibinfo
  {author} {\bibfnamefont {E.}~\bibnamefont {Rodriguez}}, \bibinfo {author}
  {\bibfnamefont {J.}~\bibnamefont {Paglione}}, \bibinfo {author}
  {\bibfnamefont {S.}~\bibnamefont {Curtarolo}}, \ and\ \bibinfo {author}
  {\bibfnamefont {I.}~\bibnamefont {Takeuchi}},\ }\href@noop {} {\bibfield
  {journal} {\bibinfo  {journal} {npj Computational Materials}\ }\textbf
  {\bibinfo {volume} {4}},\ \bibinfo {pages} {1} (\bibinfo {year}
  {2018})}\BibitemShut {NoStop}%
\bibitem [{\citenamefont {Ishikawa}\ \emph {et~al.}(2019)\citenamefont
  {Ishikawa}, \citenamefont {Miyake},\ and\ \citenamefont
  {Shimizu}}]{Ishikawa_ML_PRB2019}%
  \BibitemOpen
  \bibfield  {author} {\bibinfo {author} {\bibfnamefont {T.}~\bibnamefont
  {Ishikawa}}, \bibinfo {author} {\bibfnamefont {T.}~\bibnamefont {Miyake}}, \
  and\ \bibinfo {author} {\bibfnamefont {K.}~\bibnamefont {Shimizu}},\ }\href
  {\doibase 10.1103/PhysRevB.100.174506} {\bibfield  {journal} {\bibinfo
  {journal} {Phys. Rev. B}\ }\textbf {\bibinfo {volume} {100}},\ \bibinfo
  {pages} {174506} (\bibinfo {year} {2019})}\BibitemShut {NoStop}%
\bibitem [{\citenamefont {Xie}\ \emph {et~al.}(2019)\citenamefont {Xie},
  \citenamefont {Stewart}, \citenamefont {Hamlin}, \citenamefont {Hirschfeld},\
  and\ \citenamefont {Hennig}}]{ML:Eliashberg_Xie_PRB2019}%
  \BibitemOpen
  \bibfield  {author} {\bibinfo {author} {\bibfnamefont {S.~R.}\ \bibnamefont
  {Xie}}, \bibinfo {author} {\bibfnamefont {G.~R.}\ \bibnamefont {Stewart}},
  \bibinfo {author} {\bibfnamefont {J.~J.}\ \bibnamefont {Hamlin}}, \bibinfo
  {author} {\bibfnamefont {P.~J.}\ \bibnamefont {Hirschfeld}}, \ and\ \bibinfo
  {author} {\bibfnamefont {R.~G.}\ \bibnamefont {Hennig}},\ }\href {\doibase
  10.1103/PhysRevB.100.174513} {\bibfield  {journal} {\bibinfo  {journal}
  {Phys. Rev. B}\ }\textbf {\bibinfo {volume} {100}},\ \bibinfo {pages}
  {174513} (\bibinfo {year} {2019})}\BibitemShut {NoStop}%
\bibitem [{\citenamefont {Michael
  J.~Hutcheon}(2015)}]{ML:Hutcheon_Hydrides_arxiv2020}%
  \BibitemOpen
  \bibfield  {author} {\bibinfo {author} {\bibfnamefont {R.~J.~N.}\
  \bibnamefont {Michael J.~Hutcheon}, \bibfnamefont {Alice M.~Shipley}},\
  }\href {https://arxiv.org/abs/2001.09852} {\bibfield  {journal} {\bibinfo
  {journal} {arXiv/cond-mat/2001.09852}\ } (\bibinfo {year}
  {2015})}\BibitemShut {NoStop}%
\bibitem [{\citenamefont {Goedecker}(2004)}]{Goedecker_2004}%
  \BibitemOpen
  \bibfield  {author} {\bibinfo {author} {\bibfnamefont {S.}~\bibnamefont
  {Goedecker}},\ }\href@noop {} {\bibfield  {journal} {\bibinfo  {journal} {The
  Journal of Chemical Physics}\ }\textbf {\bibinfo {volume} {120}},\ \bibinfo
  {pages} {9911} (\bibinfo {year} {2004})}\BibitemShut {NoStop}%
\bibitem [{\citenamefont {Goedecker}\ \emph {et~al.}(2005)\citenamefont
  {Goedecker}, \citenamefont {Hellmann},\ and\ \citenamefont
  {Lenosky}}]{Goedecker_2005}%
  \BibitemOpen
  \bibfield  {author} {\bibinfo {author} {\bibfnamefont {S.}~\bibnamefont
  {Goedecker}}, \bibinfo {author} {\bibfnamefont {W.}~\bibnamefont {Hellmann}},
  \ and\ \bibinfo {author} {\bibfnamefont {T.}~\bibnamefont {Lenosky}},\
  }\href@noop {} {\bibfield  {journal} {\bibinfo  {journal} {Physical Review
  Letters}\ }\textbf {\bibinfo {volume} {95}},\ \bibinfo {pages} {055501}
  (\bibinfo {year} {2005})}\BibitemShut {NoStop}%
\bibitem [{\citenamefont {Amsler}\ and\ \citenamefont
  {Goedecker}(2010)}]{Amsler_2010}%
  \BibitemOpen
  \bibfield  {author} {\bibinfo {author} {\bibfnamefont {M.}~\bibnamefont
  {Amsler}}\ and\ \bibinfo {author} {\bibfnamefont {S.}~\bibnamefont
  {Goedecker}},\ }\href@noop {} {\bibfield  {journal} {\bibinfo  {journal} {The
  Journal of Chemical Physics}\ }\textbf {\bibinfo {volume} {133}},\ \bibinfo
  {pages} {224104} (\bibinfo {year} {2010})}\BibitemShut {NoStop}%
\bibitem [{\citenamefont {Jay}\ \emph {et~al.}(2019)\citenamefont {Jay},
  \citenamefont {Hardouin~Duparc}, \citenamefont {Sjakste},\ and\ \citenamefont
  {Vast}}]{jay2019theoretical}%
  \BibitemOpen
  \bibfield  {author} {\bibinfo {author} {\bibfnamefont {A.}~\bibnamefont
  {Jay}}, \bibinfo {author} {\bibfnamefont {O.}~\bibnamefont
  {Hardouin~Duparc}}, \bibinfo {author} {\bibfnamefont {J.}~\bibnamefont
  {Sjakste}}, \ and\ \bibinfo {author} {\bibfnamefont {N.}~\bibnamefont
  {Vast}},\ }\href@noop {} {\bibfield  {journal} {\bibinfo  {journal} {Journal
  of Applied Physics}\ }\textbf {\bibinfo {volume} {125}},\ \bibinfo {pages}
  {185902} (\bibinfo {year} {2019})}\BibitemShut {NoStop}%
\bibitem [{\citenamefont {Calandra}\ \emph {et~al.}(2004)\citenamefont
  {Calandra}, \citenamefont {Vast},\ and\ \citenamefont
  {Mauri}}]{SC:calandra_B12_2004}%
  \BibitemOpen
  \bibfield  {author} {\bibinfo {author} {\bibfnamefont {M.}~\bibnamefont
  {Calandra}}, \bibinfo {author} {\bibfnamefont {N.}~\bibnamefont {Vast}}, \
  and\ \bibinfo {author} {\bibfnamefont {F.}~\bibnamefont {Mauri}},\ }\href
  {\doibase 10.1103/PhysRevB.69.224505} {\bibfield  {journal} {\bibinfo
  {journal} {Phys. Rev. B}\ }\textbf {\bibinfo {volume} {69}},\ \bibinfo
  {pages} {224505} (\bibinfo {year} {2004})}\BibitemShut {NoStop}%
\bibitem [{\citenamefont {Calandra}\ and\ \citenamefont
  {Mauri}(2008)}]{SC:Calandra_BC5_PRL2008}%
  \BibitemOpen
  \bibfield  {author} {\bibinfo {author} {\bibfnamefont {M.}~\bibnamefont
  {Calandra}}\ and\ \bibinfo {author} {\bibfnamefont {F.}~\bibnamefont
  {Mauri}},\ }\href {\doibase 10.1103/PhysRevLett.101.016401} {\bibfield
  {journal} {\bibinfo  {journal} {Phys. Rev. Lett.}\ }\textbf {\bibinfo
  {volume} {101}},\ \bibinfo {pages} {016401} (\bibinfo {year}
  {2008})}\BibitemShut {NoStop}%
\bibitem [{\citenamefont {Moussa}\ and\ \citenamefont
  {Cohen}(2008{\natexlab{a}})}]{Moussa_diamond_PRB2008}%
  \BibitemOpen
  \bibfield  {author} {\bibinfo {author} {\bibfnamefont {J.~E.}\ \bibnamefont
  {Moussa}}\ and\ \bibinfo {author} {\bibfnamefont {M.~L.}\ \bibnamefont
  {Cohen}},\ }\href {\doibase 10.1103/PhysRevB.77.064518} {\bibfield  {journal}
  {\bibinfo  {journal} {Phys. Rev. B}\ }\textbf {\bibinfo {volume} {77}},\
  \bibinfo {pages} {064518} (\bibinfo {year} {2008}{\natexlab{a}})}\BibitemShut
  {NoStop}%
\bibitem [{\citenamefont {Moussa}\ and\ \citenamefont
  {Cohen}(2006)}]{SC:Moussa_bounds_PRB2006}%
  \BibitemOpen
  \bibfield  {author} {\bibinfo {author} {\bibfnamefont {J.~E.}\ \bibnamefont
  {Moussa}}\ and\ \bibinfo {author} {\bibfnamefont {M.~L.}\ \bibnamefont
  {Cohen}},\ }\href {\doibase 10.1103/PhysRevB.74.094520} {\bibfield  {journal}
  {\bibinfo  {journal} {Phys. Rev. B}\ }\textbf {\bibinfo {volume} {74}},\
  \bibinfo {pages} {094520} (\bibinfo {year} {2006})}\BibitemShut {NoStop}%
\bibitem [{\citenamefont {Moussa}\ and\ \citenamefont
  {Cohen}(2008{\natexlab{b}})}]{SC:Moussa_molfrag_PRB2008}%
  \BibitemOpen
  \bibfield  {author} {\bibinfo {author} {\bibfnamefont {J.~E.}\ \bibnamefont
  {Moussa}}\ and\ \bibinfo {author} {\bibfnamefont {M.~L.}\ \bibnamefont
  {Cohen}},\ }\href {\doibase 10.1103/PhysRevB.78.064502} {\bibfield  {journal}
  {\bibinfo  {journal} {Phys. Rev. B}\ }\textbf {\bibinfo {volume} {78}},\
  \bibinfo {pages} {064502} (\bibinfo {year} {2008}{\natexlab{b}})}\BibitemShut
  {NoStop}%
\bibitem [{\citenamefont {McMillan}(1968)}]{PhysRev.167.331}%
  \BibitemOpen
  \bibfield  {author} {\bibinfo {author} {\bibfnamefont {W.~L.}\ \bibnamefont
  {McMillan}},\ }\href {\doibase 10.1103/PhysRev.167.331} {\bibfield  {journal}
  {\bibinfo  {journal} {Phys. Rev.}\ }\textbf {\bibinfo {volume} {167}},\
  \bibinfo {pages} {331} (\bibinfo {year} {1968})}\BibitemShut {NoStop}%
\bibitem [{\citenamefont {Aykol}\ \emph {et~al.}(2018)\citenamefont {Aykol},
  \citenamefont {Dwaraknath}, \citenamefont {Sun},\ and\ \citenamefont
  {Persson}}]{aykol2018thermodynamic}%
  \BibitemOpen
  \bibfield  {author} {\bibinfo {author} {\bibfnamefont {M.}~\bibnamefont
  {Aykol}}, \bibinfo {author} {\bibfnamefont {S.~S.}\ \bibnamefont
  {Dwaraknath}}, \bibinfo {author} {\bibfnamefont {W.}~\bibnamefont {Sun}}, \
  and\ \bibinfo {author} {\bibfnamefont {K.~A.}\ \bibnamefont {Persson}},\
  }\href@noop {} {\bibfield  {journal} {\bibinfo  {journal} {Science advances}\
  }\textbf {\bibinfo {volume} {4}},\ \bibinfo {pages} {eaaq0148} (\bibinfo
  {year} {2018})}\BibitemShut {NoStop}%
\bibitem [{\citenamefont {Lin}\ \emph {et~al.}(2016)\citenamefont {Lin},
  \citenamefont {Zhao}, \citenamefont {Strobel},\ and\ \citenamefont
  {Cohen}}]{lin2016interpenetrating}%
  \BibitemOpen
  \bibfield  {author} {\bibinfo {author} {\bibfnamefont {Y.}~\bibnamefont
  {Lin}}, \bibinfo {author} {\bibfnamefont {Z.}~\bibnamefont {Zhao}}, \bibinfo
  {author} {\bibfnamefont {T.~A.}\ \bibnamefont {Strobel}}, \ and\ \bibinfo
  {author} {\bibfnamefont {R.}~\bibnamefont {Cohen}},\ }\href@noop {}
  {\bibfield  {journal} {\bibinfo  {journal} {Physical Review B}\ }\textbf
  {\bibinfo {volume} {94}},\ \bibinfo {pages} {245422} (\bibinfo {year}
  {2016})}\BibitemShut {NoStop}%
\bibitem [{\citenamefont {Jiang}\ \emph {et~al.}(2013)\citenamefont {Jiang},
  \citenamefont {Zhao}, \citenamefont {Li},\ and\ \citenamefont
  {Ahuja}}]{jiang2013tunable}%
  \BibitemOpen
  \bibfield  {author} {\bibinfo {author} {\bibfnamefont {X.}~\bibnamefont
  {Jiang}}, \bibinfo {author} {\bibfnamefont {J.}~\bibnamefont {Zhao}},
  \bibinfo {author} {\bibfnamefont {Y.-L.}\ \bibnamefont {Li}}, \ and\ \bibinfo
  {author} {\bibfnamefont {R.}~\bibnamefont {Ahuja}},\ }\href@noop {}
  {\bibfield  {journal} {\bibinfo  {journal} {Advanced Functional Materials}\
  }\textbf {\bibinfo {volume} {23}},\ \bibinfo {pages} {5846} (\bibinfo {year}
  {2013})}\BibitemShut {NoStop}%
\bibitem [{\citenamefont {Feng}\ \emph {et~al.}(2018)\citenamefont {Feng},
  \citenamefont {Wu}, \citenamefont {Cheng}, \citenamefont {Wen}, \citenamefont
  {Wang}, \citenamefont {Kawazoe},\ and\ \citenamefont
  {Jena}}]{feng2018monoclinic}%
  \BibitemOpen
  \bibfield  {author} {\bibinfo {author} {\bibfnamefont {X.}~\bibnamefont
  {Feng}}, \bibinfo {author} {\bibfnamefont {Q.}~\bibnamefont {Wu}}, \bibinfo
  {author} {\bibfnamefont {Y.}~\bibnamefont {Cheng}}, \bibinfo {author}
  {\bibfnamefont {B.}~\bibnamefont {Wen}}, \bibinfo {author} {\bibfnamefont
  {Q.}~\bibnamefont {Wang}}, \bibinfo {author} {\bibfnamefont {Y.}~\bibnamefont
  {Kawazoe}}, \ and\ \bibinfo {author} {\bibfnamefont {P.}~\bibnamefont
  {Jena}},\ }\href@noop {} {\bibfield  {journal} {\bibinfo  {journal} {Carbon}\
  }\textbf {\bibinfo {volume} {127}},\ \bibinfo {pages} {527} (\bibinfo {year}
  {2018})}\BibitemShut {NoStop}%
\bibitem [{\citenamefont {Chen}\ \emph {et~al.}(2015)\citenamefont {Chen},
  \citenamefont {Xie}, \citenamefont {Yang}, \citenamefont {Pan}, \citenamefont
  {Zhang}, \citenamefont {Cohen},\ and\ \citenamefont
  {Zhang}}]{chen2015nanostructured}%
  \BibitemOpen
  \bibfield  {author} {\bibinfo {author} {\bibfnamefont {Y.}~\bibnamefont
  {Chen}}, \bibinfo {author} {\bibfnamefont {Y.}~\bibnamefont {Xie}}, \bibinfo
  {author} {\bibfnamefont {S.~A.}\ \bibnamefont {Yang}}, \bibinfo {author}
  {\bibfnamefont {H.}~\bibnamefont {Pan}}, \bibinfo {author} {\bibfnamefont
  {F.}~\bibnamefont {Zhang}}, \bibinfo {author} {\bibfnamefont {M.~L.}\
  \bibnamefont {Cohen}}, \ and\ \bibinfo {author} {\bibfnamefont
  {S.}~\bibnamefont {Zhang}},\ }\href@noop {} {\bibfield  {journal} {\bibinfo
  {journal} {Nano letters}\ }\textbf {\bibinfo {volume} {15}},\ \bibinfo
  {pages} {6974} (\bibinfo {year} {2015})}\BibitemShut {NoStop}%
\bibitem [{\citenamefont {Wang}\ \emph {et~al.}(2018)\citenamefont {Wang},
  \citenamefont {Wu}, \citenamefont {Yang}, \citenamefont {Ruckenstein},\ and\
  \citenamefont {Chen}}]{wang2018semimetallic}%
  \BibitemOpen
  \bibfield  {author} {\bibinfo {author} {\bibfnamefont {S.}~\bibnamefont
  {Wang}}, \bibinfo {author} {\bibfnamefont {D.}~\bibnamefont {Wu}}, \bibinfo
  {author} {\bibfnamefont {B.}~\bibnamefont {Yang}}, \bibinfo {author}
  {\bibfnamefont {E.}~\bibnamefont {Ruckenstein}}, \ and\ \bibinfo {author}
  {\bibfnamefont {H.}~\bibnamefont {Chen}},\ }\href@noop {} {\bibfield
  {journal} {\bibinfo  {journal} {Nanoscale}\ }\textbf {\bibinfo {volume}
  {10}},\ \bibinfo {pages} {2748} (\bibinfo {year} {2018})}\BibitemShut
  {NoStop}%
\bibitem [{\citenamefont {Gao}\ \emph {et~al.}(2018)\citenamefont {Gao},
  \citenamefont {Chen}, \citenamefont {Xie}, \citenamefont {Chang},
  \citenamefont {Cohen},\ and\ \citenamefont {Zhang}}]{gao2018class}%
  \BibitemOpen
  \bibfield  {author} {\bibinfo {author} {\bibfnamefont {Y.}~\bibnamefont
  {Gao}}, \bibinfo {author} {\bibfnamefont {Y.}~\bibnamefont {Chen}}, \bibinfo
  {author} {\bibfnamefont {Y.}~\bibnamefont {Xie}}, \bibinfo {author}
  {\bibfnamefont {P.-Y.}\ \bibnamefont {Chang}}, \bibinfo {author}
  {\bibfnamefont {M.~L.}\ \bibnamefont {Cohen}}, \ and\ \bibinfo {author}
  {\bibfnamefont {S.}~\bibnamefont {Zhang}},\ }\href@noop {} {\bibfield
  {journal} {\bibinfo  {journal} {Physical Review B}\ }\textbf {\bibinfo
  {volume} {97}},\ \bibinfo {pages} {121108} (\bibinfo {year}
  {2018})}\BibitemShut {NoStop}%
\bibitem [{\citenamefont {Chen}\ \emph {et~al.}(2018)\citenamefont {Chen},
  \citenamefont {Xie}, \citenamefont {Gao}, \citenamefont {Chang},
  \citenamefont {Zhang},\ and\ \citenamefont
  {Vanderbilt}}]{PhysRevMaterials.2.044205}%
  \BibitemOpen
  \bibfield  {author} {\bibinfo {author} {\bibfnamefont {Y.}~\bibnamefont
  {Chen}}, \bibinfo {author} {\bibfnamefont {Y.}~\bibnamefont {Xie}}, \bibinfo
  {author} {\bibfnamefont {Y.}~\bibnamefont {Gao}}, \bibinfo {author}
  {\bibfnamefont {P.-Y.}\ \bibnamefont {Chang}}, \bibinfo {author}
  {\bibfnamefont {S.}~\bibnamefont {Zhang}}, \ and\ \bibinfo {author}
  {\bibfnamefont {D.}~\bibnamefont {Vanderbilt}},\ }\href {\doibase
  10.1103/PhysRevMaterials.2.044205} {\bibfield  {journal} {\bibinfo  {journal}
  {Phys. Rev. Materials}\ }\textbf {\bibinfo {volume} {2}},\ \bibinfo {pages}
  {044205} (\bibinfo {year} {2018})}\BibitemShut {NoStop}%
\bibitem [{\citenamefont {Baughman}\ and\ \citenamefont
  {Galvao}(1993)}]{baughman1993tubulanes}%
  \BibitemOpen
  \bibfield  {author} {\bibinfo {author} {\bibfnamefont {R.}~\bibnamefont
  {Baughman}}\ and\ \bibinfo {author} {\bibfnamefont {D.}~\bibnamefont
  {Galvao}},\ }\href@noop {} {\bibfield  {journal} {\bibinfo  {journal}
  {Chemical physics letters}\ }\textbf {\bibinfo {volume} {211}},\ \bibinfo
  {pages} {110} (\bibinfo {year} {1993})}\BibitemShut {NoStop}%
\bibitem [{\citenamefont {Perdew}\ and\ \citenamefont {Wang}(1992)}]{LDA-PW}%
  \BibitemOpen
  \bibfield  {author} {\bibinfo {author} {\bibfnamefont {J.~P.}\ \bibnamefont
  {Perdew}}\ and\ \bibinfo {author} {\bibfnamefont {Y.}~\bibnamefont {Wang}},\
  }\href {\doibase 10.1103/PhysRevB.45.13244} {\bibfield  {journal} {\bibinfo
  {journal} {Phys. Rev. B}\ }\textbf {\bibinfo {volume} {45}},\ \bibinfo
  {pages} {13244} (\bibinfo {year} {1992})}\BibitemShut {NoStop}%
\bibitem [{Note1()}]{Note1}%
  \BibitemOpen
  \bibinfo {note} {The threshold of distance for the presence of C-C, B-B and
  B-C bonds are d$_{CC}\le $ 1.5 \r A\protect \tmspace +\thickmuskip {.2777em},
  d$_{BB}\le $ 1.85 \r A\protect \tmspace +\thickmuskip {.2777em} and
  d$_{BC}\le $ 1.65 \r A\protect \tmspace +\thickmuskip {.2777em}
  respectively.}\BibitemShut {Stop}%
\bibitem [{\citenamefont {Zhou}\ \emph {et~al.}(2015)\citenamefont {Zhou},
  \citenamefont {Sun}, \citenamefont {Wang},\ and\ \citenamefont
  {Jena}}]{PhysRevB.92.064505}%
  \BibitemOpen
  \bibfield  {author} {\bibinfo {author} {\bibfnamefont {J.}~\bibnamefont
  {Zhou}}, \bibinfo {author} {\bibfnamefont {Q.}~\bibnamefont {Sun}}, \bibinfo
  {author} {\bibfnamefont {Q.}~\bibnamefont {Wang}}, \ and\ \bibinfo {author}
  {\bibfnamefont {P.}~\bibnamefont {Jena}},\ }\href {\doibase
  10.1103/PhysRevB.92.064505} {\bibfield  {journal} {\bibinfo  {journal} {Phys.
  Rev. B}\ }\textbf {\bibinfo {volume} {92}},\ \bibinfo {pages} {064505}
  (\bibinfo {year} {2015})}\BibitemShut {NoStop}%
\bibitem [{\citenamefont {Mann}\ \emph {et~al.}(2016)\citenamefont {Mann},
  \citenamefont {Rani}, \citenamefont {Kumar}, \citenamefont {Dubey},\ and\
  \citenamefont {Jindal}}]{mann2016thermodynamic}%
  \BibitemOpen
  \bibfield  {author} {\bibinfo {author} {\bibfnamefont {S.}~\bibnamefont
  {Mann}}, \bibinfo {author} {\bibfnamefont {P.}~\bibnamefont {Rani}}, \bibinfo
  {author} {\bibfnamefont {R.}~\bibnamefont {Kumar}}, \bibinfo {author}
  {\bibfnamefont {G.~S.}\ \bibnamefont {Dubey}}, \ and\ \bibinfo {author}
  {\bibfnamefont {V.}~\bibnamefont {Jindal}},\ }\href@noop {} {\bibfield
  {journal} {\bibinfo  {journal} {RSC advances}\ }\textbf {\bibinfo {volume}
  {6}},\ \bibinfo {pages} {12158} (\bibinfo {year} {2016})}\BibitemShut
  {NoStop}%
\bibitem [{Note2()}]{Note2}%
  \BibitemOpen
  \bibinfo {note} {The vibrational frequencies are calculated for the pure
  diamond structure with the lattice constant of D05 i.e. a=3.74 \r A\protect
  \tmspace +\thickmuskip {.2777em}, which is a $\sim 5\%$ larger than the
  experimental lattice constant of diamond.}\BibitemShut {Stop}%
\bibitem [{\citenamefont {Eliashberg}(1960)}]{eliashberg1960interactions}%
  \BibitemOpen
  \bibfield  {author} {\bibinfo {author} {\bibfnamefont {G.}~\bibnamefont
  {Eliashberg}},\ }\href@noop {} {\bibfield  {journal} {\bibinfo  {journal}
  {Sov. Phys. JETP}\ }\textbf {\bibinfo {volume} {11}},\ \bibinfo {pages} {696}
  (\bibinfo {year} {1960})}\BibitemShut {NoStop}%
\bibitem [{\citenamefont {Hopfield}(1969)}]{PhysRev.186.443}%
  \BibitemOpen
  \bibfield  {author} {\bibinfo {author} {\bibfnamefont {J.~J.}\ \bibnamefont
  {Hopfield}},\ }\href {\doibase 10.1103/PhysRev.186.443} {\bibfield  {journal}
  {\bibinfo  {journal} {Phys. Rev.}\ }\textbf {\bibinfo {volume} {186}},\
  \bibinfo {pages} {443} (\bibinfo {year} {1969})}\BibitemShut {NoStop}%
\bibitem [{\citenamefont {Amsler}(2018)}]{Book:Amsler2018}%
  \BibitemOpen
  \bibfield  {author} {\bibinfo {author} {\bibfnamefont {M.}~\bibnamefont
  {Amsler}},\ }\enquote {\bibinfo {title} {Minima hopping method for predicting
  complex structures and chemical reaction pathways},}\ in\ \href {\doibase
  10.1007/978-3-319-50257-1_77-1} {\emph {\bibinfo {booktitle} {Handbook of
  Materials Modeling: Applications: Current and Emerging Materials}}},\
  \bibinfo {editor} {edited by\ \bibinfo {editor} {\bibfnamefont
  {W.}~\bibnamefont {Andreoni}}\ and\ \bibinfo {editor} {\bibfnamefont
  {S.}~\bibnamefont {Yip}}}\ (\bibinfo  {publisher} {Springer International
  Publishing},\ \bibinfo {address} {Cham},\ \bibinfo {year} {2018})\ pp.\
  \bibinfo {pages} {1--20}\BibitemShut {NoStop}%
\bibitem [{\citenamefont {Amsler}\ \emph {et~al.}(2019)\citenamefont {Amsler},
  \citenamefont {Rostami}, \citenamefont {Tahmasbi}, \citenamefont
  {Rahmatizad}, \citenamefont {Faraji}, \citenamefont {Rasoulkhani},\ and\
  \citenamefont {Ghasemi}}]{amsler2019flame}%
  \BibitemOpen
  \bibfield  {author} {\bibinfo {author} {\bibfnamefont {M.}~\bibnamefont
  {Amsler}}, \bibinfo {author} {\bibfnamefont {S.}~\bibnamefont {Rostami}},
  \bibinfo {author} {\bibfnamefont {H.}~\bibnamefont {Tahmasbi}}, \bibinfo
  {author} {\bibfnamefont {E.}~\bibnamefont {Rahmatizad}}, \bibinfo {author}
  {\bibfnamefont {S.}~\bibnamefont {Faraji}}, \bibinfo {author} {\bibfnamefont
  {R.}~\bibnamefont {Rasoulkhani}}, \ and\ \bibinfo {author} {\bibfnamefont
  {S.~A.}\ \bibnamefont {Ghasemi}},\ }\href@noop {} {\bibfield  {journal}
  {\bibinfo  {journal} {arXiv preprint arXiv:1912.04055}\ } (\bibinfo {year}
  {2019})}\BibitemShut {NoStop}%
\bibitem [{\citenamefont {Kresse}\ and\ \citenamefont
  {J.}(1996)}]{VASP_Kresse}%
  \BibitemOpen
  \bibfield  {author} {\bibinfo {author} {\bibfnamefont {G.}~\bibnamefont
  {Kresse}}\ and\ \bibinfo {author} {\bibfnamefont {F.}~\bibnamefont {J.}},\
  }\href@noop {} {\bibfield  {journal} {\bibinfo  {journal} {Comput. Mat.
  Sci.}\ }\textbf {\bibinfo {volume} {6}},\ \bibinfo {pages} {15} (\bibinfo
  {year} {1996})}\BibitemShut {NoStop}%
\bibitem [{\citenamefont {Kresse}\ and\ \citenamefont
  {Furthm{\"u}ller}(1996)}]{kresse1996efficient}%
  \BibitemOpen
  \bibfield  {author} {\bibinfo {author} {\bibfnamefont {G.}~\bibnamefont
  {Kresse}}\ and\ \bibinfo {author} {\bibfnamefont {J.}~\bibnamefont
  {Furthm{\"u}ller}},\ }\href@noop {} {\bibfield  {journal} {\bibinfo
  {journal} {Physical review B}\ }\textbf {\bibinfo {volume} {54}},\ \bibinfo
  {pages} {11169} (\bibinfo {year} {1996})}\BibitemShut {NoStop}%
\bibitem [{\citenamefont {Kresse}\ and\ \citenamefont
  {Joubert}(1999)}]{PAW-VASP}%
  \BibitemOpen
  \bibfield  {author} {\bibinfo {author} {\bibfnamefont {G.}~\bibnamefont
  {Kresse}}\ and\ \bibinfo {author} {\bibfnamefont {D.}~\bibnamefont
  {Joubert}},\ }\href {\doibase 10.1103/PhysRevB.59.1758} {\bibfield  {journal}
  {\bibinfo  {journal} {Phys. Rev. B}\ }\textbf {\bibinfo {volume} {59}},\
  \bibinfo {pages} {1758} (\bibinfo {year} {1999})}\BibitemShut {NoStop}%
\bibitem [{\citenamefont {Perdew}\ \emph {et~al.}(1996)\citenamefont {Perdew},
  \citenamefont {Burke},\ and\ \citenamefont {Ernzerhof}}]{GGA-PBE}%
  \BibitemOpen
  \bibfield  {author} {\bibinfo {author} {\bibfnamefont {J.~P.}\ \bibnamefont
  {Perdew}}, \bibinfo {author} {\bibfnamefont {K.}~\bibnamefont {Burke}}, \
  and\ \bibinfo {author} {\bibfnamefont {M.}~\bibnamefont {Ernzerhof}},\ }\href
  {\doibase 10.1103/PhysRevLett.77.3865} {\bibfield  {journal} {\bibinfo
  {journal} {Phys. Rev. Lett.}\ }\textbf {\bibinfo {volume} {77}},\ \bibinfo
  {pages} {3865} (\bibinfo {year} {1996})}\BibitemShut {NoStop}%
\bibitem [{\citenamefont {Grimme}\ \emph {et~al.}(2011)\citenamefont {Grimme},
  \citenamefont {Ehrlich},\ and\ \citenamefont {Goerigk}}]{D3-BJ}%
  \BibitemOpen
  \bibfield  {author} {\bibinfo {author} {\bibfnamefont {S.}~\bibnamefont
  {Grimme}}, \bibinfo {author} {\bibfnamefont {S.}~\bibnamefont {Ehrlich}}, \
  and\ \bibinfo {author} {\bibfnamefont {L.}~\bibnamefont {Goerigk}},\
  }\href@noop {} {\bibfield  {journal} {\bibinfo  {journal} {Journal of
  computational chemistry}\ }\textbf {\bibinfo {volume} {32}},\ \bibinfo
  {pages} {1456} (\bibinfo {year} {2011})}\BibitemShut {NoStop}%
\bibitem [{\citenamefont {Bl\"ochl}\ \emph {et~al.}(1994)\citenamefont
  {Bl\"ochl}, \citenamefont {Jepsen},\ and\ \citenamefont
  {Andersen}}]{DFT:OKA_tetra_PRB1994}%
  \BibitemOpen
  \bibfield  {author} {\bibinfo {author} {\bibfnamefont {P.~E.}\ \bibnamefont
  {Bl\"ochl}}, \bibinfo {author} {\bibfnamefont {O.}~\bibnamefont {Jepsen}}, \
  and\ \bibinfo {author} {\bibfnamefont {O.~K.}\ \bibnamefont {Andersen}},\
  }\href {\doibase 10.1103/PhysRevB.49.16223} {\bibfield  {journal} {\bibinfo
  {journal} {Phys. Rev. B}\ }\textbf {\bibinfo {volume} {49}},\ \bibinfo
  {pages} {16223} (\bibinfo {year} {1994})}\BibitemShut {NoStop}%
\bibitem [{\citenamefont {Giannozzi}\ \emph {et~al.}(2009)\citenamefont
  {Giannozzi}, \citenamefont {Baroni}, \citenamefont {Bonini}, \citenamefont
  {Calandra}, \citenamefont {Car}, \citenamefont {Cavazzoni}, \citenamefont
  {Ceresoli}, \citenamefont {Chiarotti}, \citenamefont {Cococcioni},
  \citenamefont {Dabo}, \citenamefont {{Dal Corso}}, \citenamefont
  {de~Gironcoli}, \citenamefont {Fabris}, \citenamefont {Fratesi},
  \citenamefont {Gebauer}, \citenamefont {Gerstmann}, \citenamefont
  {Gougoussis}, \citenamefont {Kokalj}, \citenamefont {Lazzeri}, \citenamefont
  {Martin-Samos}, \citenamefont {Marzari}, \citenamefont {Mauri}, \citenamefont
  {Mazzarello}, \citenamefont {Paolini}, \citenamefont {Pasquarello},
  \citenamefont {Paulatto}, \citenamefont {Sbraccia}, \citenamefont {Scandolo},
  \citenamefont {Sclauzero}, \citenamefont {Seitsonen}, \citenamefont
  {Smogunov}, \citenamefont {Umari},\ and\ \citenamefont
  {Wentzcovitch}}]{QE-2009}%
  \BibitemOpen
  \bibfield  {author} {\bibinfo {author} {\bibfnamefont {P.}~\bibnamefont
  {Giannozzi}}, \bibinfo {author} {\bibfnamefont {S.}~\bibnamefont {Baroni}},
  \bibinfo {author} {\bibfnamefont {N.}~\bibnamefont {Bonini}}, \bibinfo
  {author} {\bibfnamefont {M.}~\bibnamefont {Calandra}}, \bibinfo {author}
  {\bibfnamefont {R.}~\bibnamefont {Car}}, \bibinfo {author} {\bibfnamefont
  {C.}~\bibnamefont {Cavazzoni}}, \bibinfo {author} {\bibfnamefont
  {D.}~\bibnamefont {Ceresoli}}, \bibinfo {author} {\bibfnamefont {G.~L.}\
  \bibnamefont {Chiarotti}}, \bibinfo {author} {\bibfnamefont {M.}~\bibnamefont
  {Cococcioni}}, \bibinfo {author} {\bibfnamefont {I.}~\bibnamefont {Dabo}},
  \bibinfo {author} {\bibfnamefont {A.}~\bibnamefont {{Dal Corso}}}, \bibinfo
  {author} {\bibfnamefont {S.}~\bibnamefont {de~Gironcoli}}, \bibinfo {author}
  {\bibfnamefont {S.}~\bibnamefont {Fabris}}, \bibinfo {author} {\bibfnamefont
  {G.}~\bibnamefont {Fratesi}}, \bibinfo {author} {\bibfnamefont
  {R.}~\bibnamefont {Gebauer}}, \bibinfo {author} {\bibfnamefont
  {U.}~\bibnamefont {Gerstmann}}, \bibinfo {author} {\bibfnamefont
  {C.}~\bibnamefont {Gougoussis}}, \bibinfo {author} {\bibfnamefont
  {A.}~\bibnamefont {Kokalj}}, \bibinfo {author} {\bibfnamefont
  {M.}~\bibnamefont {Lazzeri}}, \bibinfo {author} {\bibfnamefont
  {L.}~\bibnamefont {Martin-Samos}}, \bibinfo {author} {\bibfnamefont
  {N.}~\bibnamefont {Marzari}}, \bibinfo {author} {\bibfnamefont
  {F.}~\bibnamefont {Mauri}}, \bibinfo {author} {\bibfnamefont
  {R.}~\bibnamefont {Mazzarello}}, \bibinfo {author} {\bibfnamefont
  {S.}~\bibnamefont {Paolini}}, \bibinfo {author} {\bibfnamefont
  {A.}~\bibnamefont {Pasquarello}}, \bibinfo {author} {\bibfnamefont
  {L.}~\bibnamefont {Paulatto}}, \bibinfo {author} {\bibfnamefont
  {C.}~\bibnamefont {Sbraccia}}, \bibinfo {author} {\bibfnamefont
  {S.}~\bibnamefont {Scandolo}}, \bibinfo {author} {\bibfnamefont
  {G.}~\bibnamefont {Sclauzero}}, \bibinfo {author} {\bibfnamefont {A.~P.}\
  \bibnamefont {Seitsonen}}, \bibinfo {author} {\bibfnamefont {A.}~\bibnamefont
  {Smogunov}}, \bibinfo {author} {\bibfnamefont {P.}~\bibnamefont {Umari}}, \
  and\ \bibinfo {author} {\bibfnamefont {R.~M.}\ \bibnamefont {Wentzcovitch}},\
  }\href {http://www.quantum-espresso.org} {\bibfield  {journal} {\bibinfo
  {journal} {Journal of Physics: Condensed Matter}\ }\textbf {\bibinfo {volume}
  {21}},\ \bibinfo {pages} {395502 (19pp)} (\bibinfo {year}
  {2009})}\BibitemShut {NoStop}%
\bibitem [{\citenamefont {Giannozzi}\ \emph {et~al.}(2017)\citenamefont
  {Giannozzi}, \citenamefont {Andreussi}, \citenamefont {Brumme}, \citenamefont
  {Bunau}, \citenamefont {Nardelli}, \citenamefont {Calandra}, \citenamefont
  {Car}, \citenamefont {Cavazzoni}, \citenamefont {Ceresoli}, \citenamefont
  {Cococcioni}, \citenamefont {Colonna}, \citenamefont {Carnimeo},
  \citenamefont {Corso}, \citenamefont {de~Gironcoli}, \citenamefont {Delugas},
  \citenamefont {Jr}, \citenamefont {Ferretti}, \citenamefont {Floris},
  \citenamefont {Fratesi}, \citenamefont {Fugallo}, \citenamefont {Gebauer},
  \citenamefont {Gerstmann}, \citenamefont {Giustino}, \citenamefont {Gorni},
  \citenamefont {Jia}, \citenamefont {Kawamura}, \citenamefont {Ko},
  \citenamefont {Kokalj}, \citenamefont {K{\"u}c{\"u}kbenli}, \citenamefont
  {Lazzeri}, \citenamefont {Marsili}, \citenamefont {Marzari}, \citenamefont
  {Mauri}, \citenamefont {Nguyen}, \citenamefont {Nguyen}, \citenamefont {de-la
  Roza}, \citenamefont {Paulatto}, \citenamefont {Poncé}, \citenamefont
  {Rocca}, \citenamefont {Sabatini}, \citenamefont {Santra}, \citenamefont
  {Schlipf}, \citenamefont {Seitsonen}, \citenamefont {Smogunov}, \citenamefont
  {Timrov}, \citenamefont {Thonhauser}, \citenamefont {Umari}, \citenamefont
  {Vast}, \citenamefont {Wu},\ and\ \citenamefont {Baroni}}]{QE-2017}%
  \BibitemOpen
  \bibfield  {author} {\bibinfo {author} {\bibfnamefont {P.}~\bibnamefont
  {Giannozzi}}, \bibinfo {author} {\bibfnamefont {O.}~\bibnamefont
  {Andreussi}}, \bibinfo {author} {\bibfnamefont {T.}~\bibnamefont {Brumme}},
  \bibinfo {author} {\bibfnamefont {O.}~\bibnamefont {Bunau}}, \bibinfo
  {author} {\bibfnamefont {M.~B.}\ \bibnamefont {Nardelli}}, \bibinfo {author}
  {\bibfnamefont {M.}~\bibnamefont {Calandra}}, \bibinfo {author}
  {\bibfnamefont {R.}~\bibnamefont {Car}}, \bibinfo {author} {\bibfnamefont
  {C.}~\bibnamefont {Cavazzoni}}, \bibinfo {author} {\bibfnamefont
  {D.}~\bibnamefont {Ceresoli}}, \bibinfo {author} {\bibfnamefont
  {M.}~\bibnamefont {Cococcioni}}, \bibinfo {author} {\bibfnamefont
  {N.}~\bibnamefont {Colonna}}, \bibinfo {author} {\bibfnamefont
  {I.}~\bibnamefont {Carnimeo}}, \bibinfo {author} {\bibfnamefont {A.~D.}\
  \bibnamefont {Corso}}, \bibinfo {author} {\bibfnamefont {S.}~\bibnamefont
  {de~Gironcoli}}, \bibinfo {author} {\bibfnamefont {P.}~\bibnamefont
  {Delugas}}, \bibinfo {author} {\bibfnamefont {R.~A.~D.}\ \bibnamefont {Jr}},
  \bibinfo {author} {\bibfnamefont {A.}~\bibnamefont {Ferretti}}, \bibinfo
  {author} {\bibfnamefont {A.}~\bibnamefont {Floris}}, \bibinfo {author}
  {\bibfnamefont {G.}~\bibnamefont {Fratesi}}, \bibinfo {author} {\bibfnamefont
  {G.}~\bibnamefont {Fugallo}}, \bibinfo {author} {\bibfnamefont
  {R.}~\bibnamefont {Gebauer}}, \bibinfo {author} {\bibfnamefont
  {U.}~\bibnamefont {Gerstmann}}, \bibinfo {author} {\bibfnamefont
  {F.}~\bibnamefont {Giustino}}, \bibinfo {author} {\bibfnamefont
  {T.}~\bibnamefont {Gorni}}, \bibinfo {author} {\bibfnamefont
  {J.}~\bibnamefont {Jia}}, \bibinfo {author} {\bibfnamefont {M.}~\bibnamefont
  {Kawamura}}, \bibinfo {author} {\bibfnamefont {H.-Y.}\ \bibnamefont {Ko}},
  \bibinfo {author} {\bibfnamefont {A.}~\bibnamefont {Kokalj}}, \bibinfo
  {author} {\bibfnamefont {E.}~\bibnamefont {K{\"u}c{\"u}kbenli}}, \bibinfo
  {author} {\bibfnamefont {M.}~\bibnamefont {Lazzeri}}, \bibinfo {author}
  {\bibfnamefont {M.}~\bibnamefont {Marsili}}, \bibinfo {author} {\bibfnamefont
  {N.}~\bibnamefont {Marzari}}, \bibinfo {author} {\bibfnamefont
  {F.}~\bibnamefont {Mauri}}, \bibinfo {author} {\bibfnamefont {N.~L.}\
  \bibnamefont {Nguyen}}, \bibinfo {author} {\bibfnamefont {H.-V.}\
  \bibnamefont {Nguyen}}, \bibinfo {author} {\bibfnamefont {A.~O.}\
  \bibnamefont {de-la Roza}}, \bibinfo {author} {\bibfnamefont
  {L.}~\bibnamefont {Paulatto}}, \bibinfo {author} {\bibfnamefont
  {S.}~\bibnamefont {Poncé}}, \bibinfo {author} {\bibfnamefont
  {D.}~\bibnamefont {Rocca}}, \bibinfo {author} {\bibfnamefont
  {R.}~\bibnamefont {Sabatini}}, \bibinfo {author} {\bibfnamefont
  {B.}~\bibnamefont {Santra}}, \bibinfo {author} {\bibfnamefont
  {M.}~\bibnamefont {Schlipf}}, \bibinfo {author} {\bibfnamefont {A.~P.}\
  \bibnamefont {Seitsonen}}, \bibinfo {author} {\bibfnamefont {A.}~\bibnamefont
  {Smogunov}}, \bibinfo {author} {\bibfnamefont {I.}~\bibnamefont {Timrov}},
  \bibinfo {author} {\bibfnamefont {T.}~\bibnamefont {Thonhauser}}, \bibinfo
  {author} {\bibfnamefont {P.}~\bibnamefont {Umari}}, \bibinfo {author}
  {\bibfnamefont {N.}~\bibnamefont {Vast}}, \bibinfo {author} {\bibfnamefont
  {X.}~\bibnamefont {Wu}}, \ and\ \bibinfo {author} {\bibfnamefont
  {S.}~\bibnamefont {Baroni}},\ }\href
  {http://stacks.iop.org/0953-8984/29/i=46/a=465901} {\bibfield  {journal}
  {\bibinfo  {journal} {Journal of Physics: Condensed Matter}\ }\textbf
  {\bibinfo {volume} {29}},\ \bibinfo {pages} {465901} (\bibinfo {year}
  {2017})}\BibitemShut {NoStop}%
\bibitem [{\citenamefont {Hamann}(2013)}]{hamann2013optimized}%
  \BibitemOpen
  \bibfield  {author} {\bibinfo {author} {\bibfnamefont {D.}~\bibnamefont
  {Hamann}},\ }\href@noop {} {\bibfield  {journal} {\bibinfo  {journal}
  {Physical Review B}\ }\textbf {\bibinfo {volume} {88}},\ \bibinfo {pages}
  {085117} (\bibinfo {year} {2013})}\BibitemShut {NoStop}%
\bibitem [{\citenamefont {Van~Setten}\ \emph {et~al.}(2018)\citenamefont
  {Van~Setten}, \citenamefont {Giantomassi}, \citenamefont {Bousquet},
  \citenamefont {Verstraete}, \citenamefont {Hamann}, \citenamefont {Gonze},\
  and\ \citenamefont {Rignanese}}]{van2018pseudodojo}%
  \BibitemOpen
  \bibfield  {author} {\bibinfo {author} {\bibfnamefont {M.}~\bibnamefont
  {Van~Setten}}, \bibinfo {author} {\bibfnamefont {M.}~\bibnamefont
  {Giantomassi}}, \bibinfo {author} {\bibfnamefont {E.}~\bibnamefont
  {Bousquet}}, \bibinfo {author} {\bibfnamefont {M.~J.}\ \bibnamefont
  {Verstraete}}, \bibinfo {author} {\bibfnamefont {D.~R.}\ \bibnamefont
  {Hamann}}, \bibinfo {author} {\bibfnamefont {X.}~\bibnamefont {Gonze}}, \
  and\ \bibinfo {author} {\bibfnamefont {G.-M.}\ \bibnamefont {Rignanese}},\
  }\href@noop {} {\bibfield  {journal} {\bibinfo  {journal} {Computer Physics
  Communications}\ }\textbf {\bibinfo {volume} {226}},\ \bibinfo {pages} {39}
  (\bibinfo {year} {2018})}\BibitemShut {NoStop}%
\end{thebibliography}%

\end{document}